\documentclass[11pt]{article}
\usepackage{amsmath,amssymb,mathrsfs,bbm,bm}
\usepackage{enumitem,bbm}
\usepackage{booktabs}
\usepackage{subcaption}
\usepackage{upgreek}
\usepackage{bigints}
\usepackage{colonequals}
\usepackage[official]{eurosym}
\usepackage[margin=1in]{geometry}
\usepackage[onehalfspacing]{setspace}
\usepackage{color}
\usepackage{graphicx,subcaption}
\usepackage{comment}
\usepackage{rotating}
\usepackage{hyperref}
\usepackage{multirow}
\usepackage{threeparttable}
\hypersetup{pdfborder = {0 0 0},colorlinks=true,linkcolor=blue,urlcolor=blue,citecolor=blue}

\usepackage[round]{natbib}

\usepackage{dutchcal} 
\usepackage[mathscr]{euscript} 

\renewcommand{\P}{\mathbbm{P}}
\newcommand{\E}{\mathbbm{E}}
\newcommand{\W}{\mathcal{W}}

\newcommand{\I}{\mathbbm{1}}

\newcommand{\C}{c}
\newcommand{\D}{D}
\newcommand{\T}{T}

\begin{document}
	
\title{A Guide to Regression Discontinuity Designs in Medical Applications\thanks{We thank our current and former collaborators Sebastian Calonico, Max Farrell, Yingjie Feng, Brigham Frandsen, Nicolas Idrobo, Michael Jansson, Xinwei Ma, Kenichi Nagasawa, Filippo Palomba, Jasjeet Sekhon, Gonzalo Vazquez-Bare, and Rae Yu for their intellectual input to our research program on RD designs. We also thank the co-Editor-in-Chief, Lisa McShane, an associate editor, and three reviewers for helpful comments. Cattaneo and Titiunik gratefully acknowledge financial support from the National Science Foundation (SES-2019432 and SES-2241575), and Cattaneo gratefully acknowledges financial support from the National Institute of Health (R01 GM072611-16).
}}

\author{
	Matias D. Cattaneo\footnote{Department of Operations Research and Financial Engineering, Princeton University.}\and
	Luke Keele\footnote{Department of Surgery and Biostatistics, University of Pennsylvania.}\and
	Roc\'{i}o Titiunik\footnote{Department of Politics, Princeton University.}}

\maketitle
\setcounter{page}{0}\thispagestyle{empty}

\begin{abstract}
    We present a practical guide for the analysis of regression discontinuity (RD) designs in biomedical contexts. We begin by introducing key concepts, assumptions, and estimands within both the continuity-based framework and the local randomization framework. We then discuss modern estimation and inference methods within both frameworks, including approaches for bandwidth or local neighborhood selection, optimal treatment effect point estimation, and robust bias-corrected inference methods for uncertainty quantification. We also overview empirical falsification tests that can be used to support key assumptions. Our discussion focuses on two particular features that are relevant in biomedical research: (i) fuzzy RD designs, which often arise when therapeutic treatments are based on clinical guidelines but patients with scores near the cutoff are treated contrary to the assignment rule; and (ii) RD designs with discrete scores, which are ubiquitous in biomedical applications. We illustrate our discussion with three empirical applications: the effect of CD4 guidelines for anti-retroviral therapy on retention of HIV patients in South Africa, the effect of genetic guidelines for chemotherapy on breast cancer recurrence in the United States, and the effects of age-based patient cost-sharing on healthcare utilization in Taiwan. We provide replication materials employing publicly available statistical software in \texttt{Python}, \texttt{R} and \texttt{Stata}, offering researchers all necessary tools to conduct an RD analysis.
\end{abstract}


\setcounter{page}{0}
\thispagestyle{empty}

\clearpage
\setcounter{page}{0}
\thispagestyle{empty}
\setcounter{tocdepth}{2}
\tableofcontents
\clearpage

\doublespacing\newpage
\section{Introduction}

Drawing causal inferences from quantitative data is a fundamental goal in epidemiology, comparative effectiveness, health services, and outcomes research \citep{Craig-Katikireddi-Leyland-Popham_2017_ARPH,Hernan_2018_AJPH,Hernan-Robins_2022_Book}. It is now well understood that while randomized controlled trials are the gold standard for learning about treatment effects, reliance on  observational studies is unavoidable---there are simply too many contexts where randomization is infeasible or unethical. When randomization is not possible, evidence from natural experiments is often viewed as the next best alternative for causal inference and program evaluation \citep{Rosenbaum_2010_Book,Imbens-Rubin_2015_Book,Abadie-Cattaneo_2018_ARE,Hernan-Robins_2022_Book}. Some scholars have advocated for greater use of the regression discontinuity (RD) design in biomedical contexts \citep{Bor-Moscoe-Mutevedzi-Newell-Barnighausen_2014_Epidem,OKeeffe-Geneletti-Baio-sharples-Nazareth-Petersen_2014_BMJ,Bor-Moscoe-Barnighausen_2015_Epidem,Maciejewski-Basu_2020_JAMA}, which can be viewed as a prime example of a natural experiment \citep{Titiunik_2021_HandbookCh}. As a result, RD designs have become more common in biomedical research: a recent review identified over $325$ studies based on RD designs in medical studies alone \citep{boon2021regression}.

The popularity of the RD design stems from its high internal validity. Causal inferences from RD designs are often more credible and robust than those from other non-experimental impact evaluation strategies such as selection-on-observables, difference-in-difference, or instrumental variable (IV) designs. The feature that contributes to the superior credibility of the RD design is the existence of an objective and verifiable treatment assignment rule that offers a design-based way to validate some of its key assumptions. In the canonical RD design, each unit $i$ receives a score $X_i$, and a treatment is assigned according to the rule $T_i=\I(X_i\geq c)$, where $c$ is a fixed known cutoff and $\I(\cdot)$ the indicator function, so that all units with score above the cutoff are assigned to the active treatment condition ($T_i=1$) and all units with scores below the cutoff are assigned to the control condition ($T_i=0$). The score $X_i$ can be continuous (each unit has a unique score value) or discrete (multiple units share the same score value). In the so-called \textit{sharp} RD design, all units comply perfectly with the treatment condition they are assigned: no units below the cutoff receive the treatment and no units above the cutoff refuse the treatment. In the more general case, referred to as the \textit{fuzzy} RD design, the treatment assignment rule induces many units to take the treatment, but compliance with the assignment is imperfect. Each variant of the RD design requires conceptually and methodologically different approaches for analysis.

The RD design was first introduced by \citet{Thistlethwaite-Campbell_1960_JEP} in education research to study the effect of receiving a certificate of merit  based on test scores. In biomedical contexts, RD designs arise naturally from treatment guidelines based on diagnostic test results. For instance, a specific treatment might be recommended when test results exceed a known cutoff---e.g. start blood pressure medication when systolic blood pressure is above $130$ mmHg. The key idea behind the RD design is that patients just above and just below the cutoff should be comparable in terms of all unobservable and observable characteristics not affected by the treatment, which in turn implies that these patients' differences in outcomes can be understood as the result of differences in treatment status rather than the result of differences in observable or unobservable characteristics. For example, assuming that patients do not have precise control over their blood pressure measurement, patients whose systolic pressure is $130$ mmHg should be similar to patients whose systolic pressure is $129$ mmHg: in a small neighborhood around the $130$ cutoff, patients' particular measures will be governed by random chance (variable device accuracy, inadequate arm support, elevated anxiety, etc.) more than by patients' underlying health risks or other confounding factors affecting the outcome of interest.

We provide a systematic overview of the state-of-the-art statistical methodologies to analyze and interpret RD designs employing the two most widely used methodological frameworks: the continuity-based framework and the local randomization framework. Our discussion covers key assumptions, estimation methods, inference procedures, and diagnostic tests complementing and expanding on early introductory articles for the biomedical sciences \citep{Bor-Moscoe-Mutevedzi-Newell-Barnighausen_2014_Epidem,OKeeffe-Geneletti-Baio-sharples-Nazareth-Petersen_2014_BMJ,Maciejewski-Basu_2020_JAMA}, which do not discuss the most recent RD methods that are now widely used in the statistical, social, and behavioral sciences \citep{Cattaneo-Titiunik_2022_ARE,Cattaneo-Idrobo-Titiunik_2020_Book,Cattaneo-Idrobo-Titiunik_2023_Book}. Furthermore, we discuss two complications that frequently arise in biomedical research that have not been addressed by prior biomedical reviews: imperfect treatment compliance and non-continuous score variables.

The manuscript is organized as follows. Sections \ref{sec:RDgeneral} and \ref{sec:Analysis with Continuous Score} focus on RD methodology when the score is (approximately) continuous; for illustration, we reanalyze and expand a recent study that used the RD design to estimate the effect of immediate versus deferred anti-retroviral therapy (ART) on retention in care \citep{Bor-Fox-Rosen-etal_2017_PloSMed}. In this application, the score (CD4 count) takes on many distinct values and thus can be analyzed using RD methods suitable for continuous scores. Section \ref{sec:RD Analysis with Discrete Score} then discusses RD designs with a discrete score, that is, settings where the score takes on at most a few distinct values. We overview how the methods for RD designs with a (approximately) continuous score can be modified and extended, illustrating our discussion with two additional empirical examples. One example looks at genetic guidelines for chemotherapy and serves mostly as a cautionary tale because the key RD assumptions are not supported empirically. The other example studies patient cost-sharing and healthcare utilization and showcases how RD methods with discrete scores can be deployed successfully. Finally, Section \ref{sec:conclusion} summarizes key takeaways for practice and concludes. The online materials include the three data sets as well as computer code to replicate all our analyses. Replication codes are available in \texttt{Python}, \texttt{R}, and \texttt{Stata}, and can be found at \url{https://rdpackages.github.io/}. The supplementary materials also include code to demonstrate a basic RD analysis.

\section{Setup and Treatment Effects}
\label{sec:RDgeneral}

In the canonical RD design, there are $i=1,2,\ldots,n$ units of analysis, each unit receives a \textit{score} $X_i$ (also known as \textit{running variable}, \textit{forcing variable}, or \textit{index}), and a binary treatment is assigned based on whether this score exceeds or not a known cutoff $\C$: units whose score is above the cutoff are assigned to the treatment condition, and units whose score is below the cutoff are assigned to the control condition. Thus, the probability of treatment assignment as a function of the score changes discontinuously at the cutoff: all units above the cutoff are assigned to the treatment condition with probability one, while all units below the cutoff are assigned to the control condition with probability one. These three elements---score, cutoff, and treatment---are the key components of all RD designs. Crucially, the RD treatment assignment rule is known, at least to the researcher, and hence empirically verifiable. This distinctive feature contributes to the RD design's superior credibility when compared to other non-experimental methods.

\subsection{Empirical Example: ART and Retention in Care}

We revisit the recent study by \citet{Bor-Fox-Rosen-etal_2017_PloSMed}, who used a RD design to estimate the effect of immediate (versus deferred) anti-retroviral therapy (ART) on retention in care. The authors analyzed the Hlabisa HIV Treatment and Care Programme in South Africa, conducted by the Africa Health Research Institute and the South African Department of Health. This program collected data on all patients receiving HIV care and treatment services at government facilities ($17$ clinics and $1$ hospital) between $12$ August $2011$ and $31$ December $2012$ \citep{Tanser-etal_2007_IJE,Houlihan-etal_2010_IJE}. Patients were eligible for ART if their CD4 count was less than 350 cells/$\upmu$l, and they had a WHO stage III/IV condition. Patients did an initial blood draw for a CD4 count, and were instructed to return to the clinic in one week to receive their result. ART-eligible patients were enrolled in several weeks of counseling and were then initiated on ART. 

The investigators compared differences in retention between patients with CD4 counts ($X_i$) just above versus just below the 350-cells/$\upmu$l threshold ($c$). The cohort included $n=11,306$ patients and the data contained information on several predetermined covariates, including sex, age, date of testing, and testing location. This is a prototypical biomedical RD example, where the units of analysis are patients and the score (CD4 count) is the result of a diagnostic test. The cutoff is 350 cells/$\upmu$l and the treatment is the immediate initiation of ART. The outcome of interest is a binary variable with value $1$ if there was any evidence of any routine clinic visits, lab result (CD4 or viral load), or date of ART initiation $6$ to $18$ months after a patient’s first CD4 count, regardless of receipt of ART. The RD design is fuzzy because not all patients with a score of less than $350$ initiated ART (see Figure \ref{fig: ART - T vs D} in the next section). Henceforth, we refer to this example as the \textit{ART} application.

Note that the treatment is assigned when the CD4 count is below (rather than above) the 350 cutoff. However, the score and cutoff can always be redefined so that treatment assignment occurs when $X_i\geq c$ (i.e., multiplying both variables by $-1$, a relabelling of the data with no substantive effects in the analysis). Alternatively, the analysis can proceed given the setup and treatment effects are simply understood with a change of sign.

\subsection{Graphical Illustration of RD Design}

\begin{figure}[ht]
    \centering
	\begin{subfigure}{0.48\textwidth}
		\centering
		\includegraphics[scale=0.45]{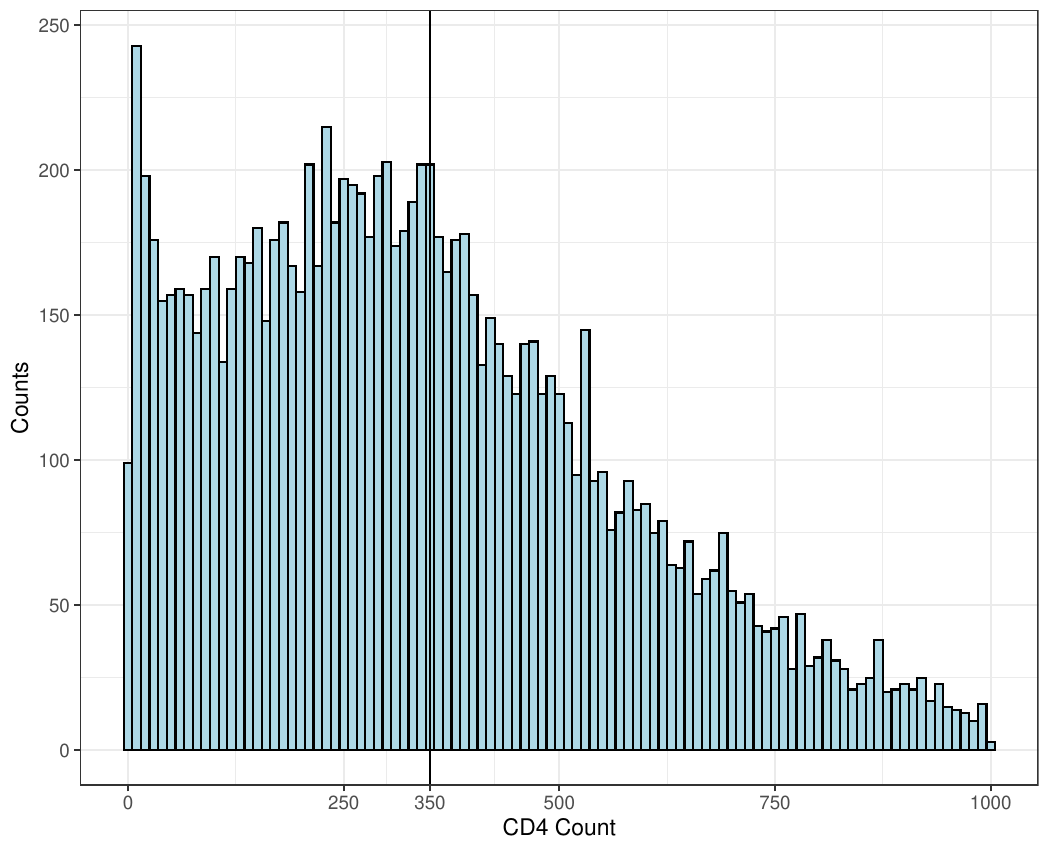}
		\caption{Histogram: Score ($X_i$)}\label{fig: ART - Histogram X}
	\end{subfigure}
	\begin{subfigure}{0.48\textwidth}
		\centering
		\includegraphics[scale=0.45]{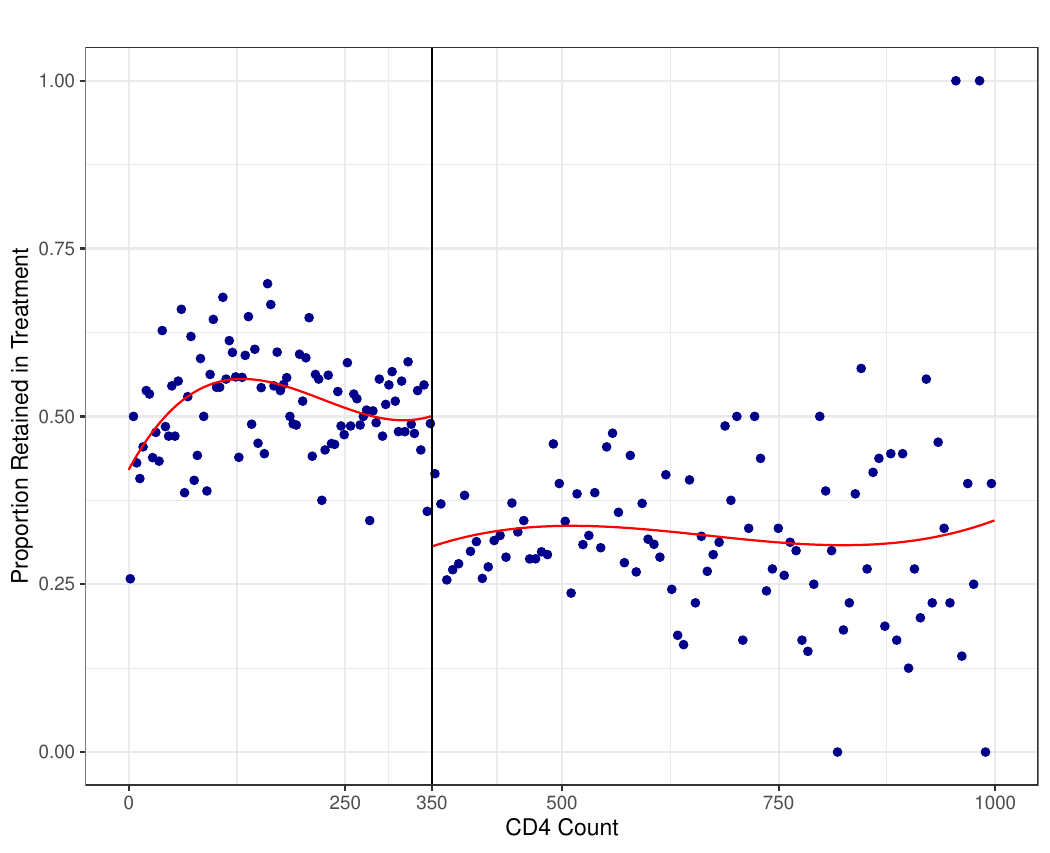}
		\caption{RD Plot: Outcome ($Y_i$) vs. Score ($X_i$)}\label{fig: ART - RD plot Y-X}		
	\end{subfigure}
	\caption{Basic Plots --- ART Application}\label{fig: ART - Basic plots}
  \flushleft{{\footnotesize Note: The score $X_i$ is patient $i$'s CD4 count, and all patients below the cutoff are assigned to receive ART. In panel (b), the outcome is $Y_i$, an indicator equal to 1 if patient $i$ was retained in care (0 otherwise); dots are local means of $Y_i$ calculated for patients in different non-overlapping bins of $X_i$; and the solid line is a 4th-order polynomial of $Y_i$ on $X_i$, fitted separately for patients above and below the cutoff.}}
\end{figure}

An important first step in RD analysis is a graphical illustration of the design. Figure \ref{fig: ART - Basic plots} showcases two basic plots. When properly executed, a graphical RD analysis adds transparency and credibility by displaying the observations used for estimation and inference, both globally (over the entire support of the score) and locally (near the cutoff determining treatment assignment). RD plots can also highlight other features of the design such as the coarseness of the score and outcome variables, the variability of the data, and the potential curvature of the underlying regression functions \citep{Calonico-Cattaneo-Titiunik_2015_JASA}. Despite their visual usefulness, RD plots should not be used as the main tool for the analysis, as they can often be misleading \citep{Korting-etal_2023_visual}; their main role should be as supplementary to the formal statistical analyses that we discuss in Section \ref{sec:Analysis with Continuous Score}.

Figure \ref{fig: ART - Histogram X} depicts a histogram of the score variable $X_i$, which captures the relative frequency of different observed values by first binning its support. In general, the score can be continuously or discretely distributed. When the score is continuous each unit has a unique score value, while when the score is discrete several units share the same score value and thus the score exhibits ``mass points.'' In the ART application, $X_i$ takes on $K=1,229$ distinct values in a sample of 11,306 total observations, so several observations share the same score value ($K < n=11,306$). Given the sizable number of distinct values, we treat this application as having an approximately continuous score and discuss RD methods for that context. This is a common approach in practice when the discrete score has ``many'' mass points \citep{Cattaneo-Titiunik_2022_ARE,Cattaneo-Idrobo-Titiunik_2023_Book}, and was the approach adopted by \citet{Bor-Fox-Rosen-etal_2017_PloSMed}. In Section \ref{sec:RD Analysis with Discrete Score}, we discuss RD methods appropriate in cases when the score is discrete with possibly only a ``few'' distinct values. We also use Figure \ref{fig: ART - Histogram X} as the starting point for validation of the RD design via discontinuity-in-density testing \citep{McCrary_2008_JoE} in Section \ref{sec:Analysis with Continuous Score}.

Figure \ref{fig: ART - RD plot Y-X} presents a canonical RD plot of the observed outcome variable given the score \citep{Calonico-Cattaneo-Titiunik_2015_JASA}. Although we could construct a raw scatter plot of the outcome against the score, such plot is often uninformative and hides many interesting features of the outcome-score relationship like discontinuities or non-linearities. For this reason, it is customary to ``smooth'' the data before plotting, which is done by binning the support of the score into disjoint (i.e., non-overlapping) intervals, and then reporting the average outcome for units with score within each bin, an approach conceptually analogous to the histogram in Figure \ref{fig: ART - Histogram X}. These binned means can be interpreted as a non-smooth local approximation to the unknown regression functions of the outcome $Y_i$ given $X_i$. The standard RD plot consists of these binned means with the addition of two global polynomials fits, one above and one below the cutoff, based on regressing $Y_i$ on a polynomial of $X_i$ using the raw (i.e., not binned) data. The global polynomial fits can be interpreted as a smooth global approximation of the unknown regression functions, in contrast to the non-smooth approximation provided by the local means. Choosing the appropriate global polynomial order is important: when the order of the polynomial is ``too'' high, the global polynomial regression will over-fit the data. This over-fitting is usually referred to as Runge's phenomenon, and is known to be particularly detrimental at boundary points, which is the area of interest in RD designs. See \citet[Section 2]{Cattaneo-Idrobo-Titiunik_2020_Book} for more details, and \citet{Cattaneo-Crump-Farrell-Feng_2023_Binscatter} for related visualization methods in other empirical contexts. 

The RD plot in Figure \ref{fig: ART - RD plot Y-X} gives a first glance at the RD design in the ART application. It shows that patients assigned to treatment (CD4 count strictly below 350) had an average retention in care higher than those assigned to control. The question is whether this difference can be interpreted as the causal effect of assignment to ART. Because the RD design focuses on a treatment effect that occurs at or near the cutoff, this interpretation requires assuming that, at the cutoff, the treatment assignment changes discontinuously but all other confounders change smoothly or not at all. To formalize this intuition, we need to introduce key assumptions underlying RD designs and also define treatment effects (or parameters) of interest by ``localizing'' near the cutoff. Following the taxonomy introduced by \citet{Cattaneo-Titiunik-VazquezBare_2017_JPAM}, we consider the continuity-based and the local randomization frameworks for the analysis and interpretation of both sharp and fuzzy RD designs.

\subsection{Sharp RD Designs}

We first discuss settings with perfect treatment compliance. This is not the case for the ART application, or many other biomedical applications, but this simpler setup helps us put forth key concepts without added complications. The next section generalizes the setup to allow for imperfect compliance (i.e., fuzzy RD designs). As discussed there, if researchers are interested on intention-to-treat effects, then the sharp RD design is indeed the appropriate setup to consider even in the presence of non-compliance. As a result, the discussion in this section is a key building block for Fuzzy RD analysis. 

We adopt the standard potential outcomes framework and assume that each unit has one outcome corresponding to each possible value of the treatment assignment: $Y_i(0)$ under control assignment, and $Y_i(1)$ under treatment assignment. The observed outcome is determined by the potential outcome corresponding to the treatment assigned to each unit: $Y_i = (1-\T_i) \cdot Y_i(0) + \T_i \cdot Y_i(1)$. The observed data is $(Y_1,X_1),\ldots,(Y_n,X_n)$. In most of our discussion, we assume that the observations are a random sample with random potential outcomes. We deviate from this setup only when discussing analysis of experiments approaches based on Fisherian inference or Neyman methods, which assume that the potential outcomes are non-stochastic and hence that the observed outcomes are random only because of the randomness induced by the treatment assignment mechanism, that is, the probability distribution determining $(T_1,T_2,\dots,T_n)$.

\subsubsection{Continuity-Based Framework}

In the sharp RD design with continuous score, the leading conceptual approach is the continuity-based framework \citep{Hahn-Todd-vanderKlaauw_2001_ECMA}, where the causal treatment effect is the average treatment effect at the cutoff:
\begin{equation}
    \tau_{\mathtt{SRD}} \equiv \E[Y_i(1) - Y_i(0) | X_i=\C].
\end{equation}
In this framework, potential outcomes are always assumed to be random, so the conditional expectations are interpreted and computed in the usual way. The sharp RD treatment effect $\tau_{\mathtt{SRD}}$ is the average effect of treatment for units \textit{local} to the cutoff---that is, for units with score values $X_i=\C$.  

The identification of $\tau_{\mathtt{SRD}}$ is based on the idea that units with similar values of the score but on opposite sides of the cutoff should be ``comparable'' in all predetermined characteristics except for the fact that units whose scores are above the cutoff are assigned to treatment while units whose scores are below the cutoff are not. Predetermined characteristics, also known as pre-treatment or predetermined covariates, are all features of the units whose values are determined before the treatment is assigned. For example, in the ART example, the age and sex of patients are predetermined covariates.

We can formalize the logic of comparability at the cutoff using continuity, which relies on mild extrapolation for units with score near the cutoff. First, we define the average potential outcomes given the score: $\E[Y_i(1) | X_i=x]$ and $\E[Y_i(0)|X_i=x]$. These conditional expectation functions are usually called \textit{regression functions}, and are unknown; the two solid lines in Figure \ref{fig: ART - RD plot Y-X} depict global polynomial approximations to these functions in the ART application. If the regression functions $\E[Y_i(1) | X_i=x]$ and $\E[Y_i(0) | X_i=x]$, seen as functions of $x$, are continuous at $x=\C$, then the units will be comparable ``just'' above and below the cutoff. That is, under the assumption of continuity, we can use the regression functions to link observed data to counterfactual quantities in the following way:
\begin{equation}\label{eq:SRD}
\tau_{\mathtt{SRD}} = \lim_{x\downarrow{\C}} \E[Y_i|X_i=x] - \lim_{x\uparrow{\C}} \E[Y_i|X_i=x].
\end{equation}

In Equation \eqref{eq:SRD}, continuity implies that as the score value gets closer to the cutoff $\C$, the average potential outcome function $\E[Y_i(0) | X_i=x]$ gets closer to its value at the cutoff, $\E[Y_i(0) | X_i=\C]$, and analogously for $\E[Y_i(1) | X_i=x]$. Thus, continuity gives a formal justification for estimating the sharp RD effect by focusing on observations in a small neighborhood above and below the cutoff to estimate, respectively and separately, $\E[Y_i(1) | X_i=\C]$  and $\E[Y_i(0) | X_i=\C]$. The observations in this neighborhood, by construction, will have similar score values; and by virtue of continuity, their average potential outcomes will also be similar. As mentioned above, employing global polynomial approximations should be avoided due to poor boundary behavior of such estimates; instead, the approximation should be local.

\begin{figure}[ht]
    \centering
	\begin{subfigure}{0.48\textwidth}
		\centering
		\includegraphics[scale=0.45]{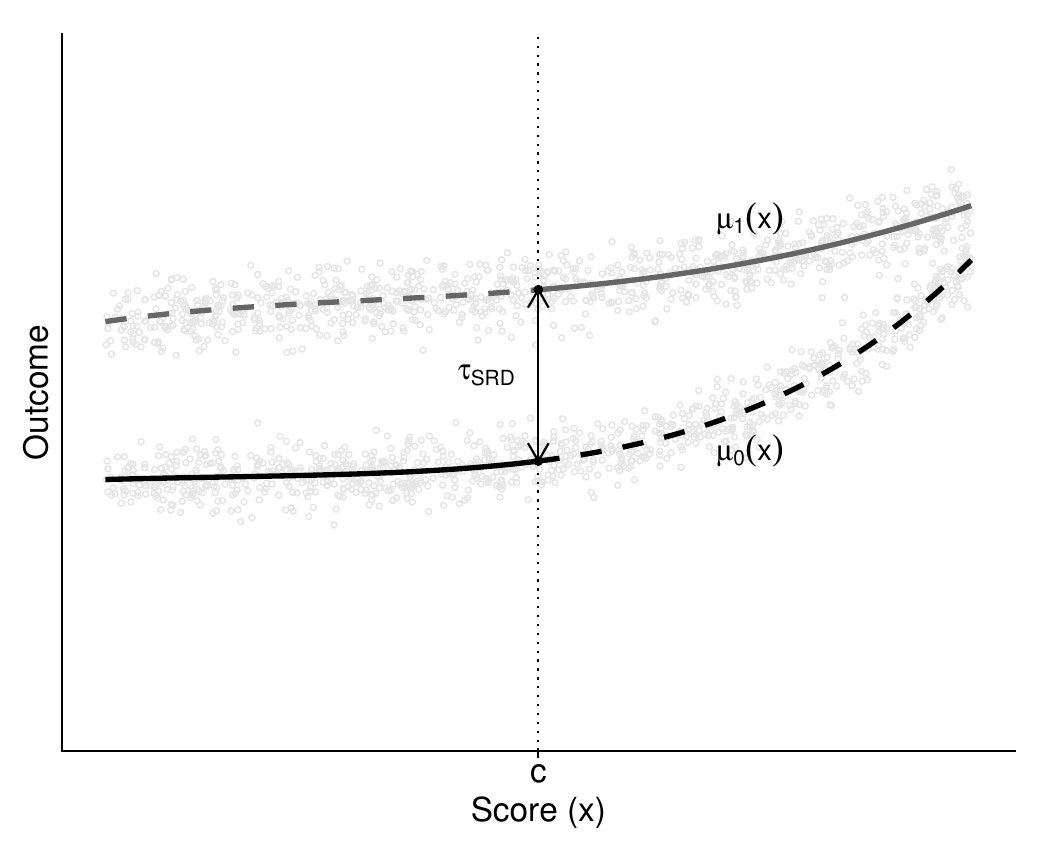}
		\caption{Continuity-Based Framework}\label{fig:Continuity-Based}
	\end{subfigure}
	\begin{subfigure}{0.48\textwidth}
		\centering
		\includegraphics[scale=0.45]{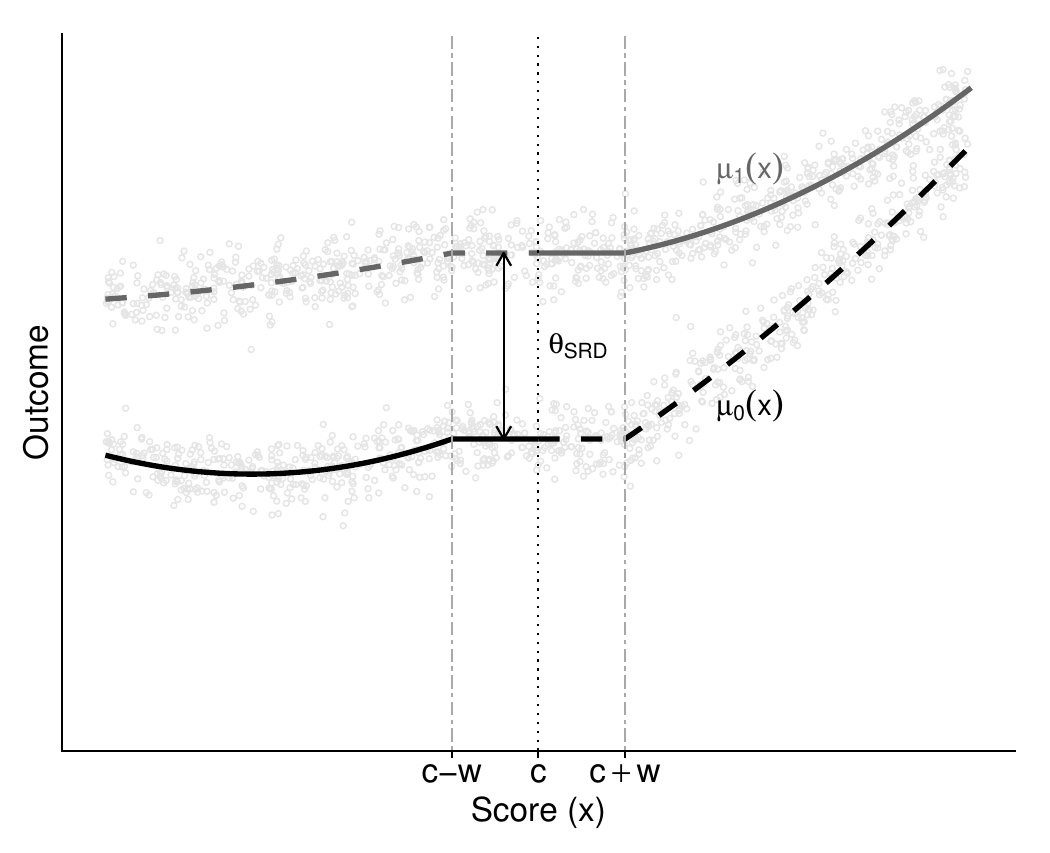}
		\caption{Local Randomization Framework}\label{fig:Local Randomization}		
	\end{subfigure}
	\caption{Graphical illustration of sharp RD design frameworks}\label{fig:RD Frameworks}
 \flushleft{{\footnotesize Note: $\mu_1(x)$ and $\mu_0(x)$ are the conditional expectation functions of the potential outcome under treatment and control, respectively, given the score---i.e., $\mu_1(x) = \E[Y_i(1) | X_i=x]$ and  $\mu_0(x) = \E[Y_i(0) | X_i=x]$. Dashed lines represent unobserved functions, solid lines represent observed functions.}}
\end{figure}

The logic of the continuity-based framework is graphically illustrated in Figure \ref{fig:Continuity-Based}. Continuity of the two conditional expectations ensures that the vertical distance between the two curves at $\C$ represents the RD estimand. We cannot directly estimate this quantity since we never observe the two curves at $\C$: units with scores exactly at or just above $\C$ are treated, but units with scores just below $\C$ are control. Nevertheless, if the average potential outcomes at $\C$ are not abruptly different from the average potential outcomes at values of the score just below $\C$, then units just above and below the cutoff should be comparable, and we can approximately identify the vertical distance at $\C$ using the local observed data, relying on minimal extrapolation in finite samples. 

The RD treatment effect $\tau_{\mathtt{SRD}}$ differs from the two most common estimands often targeted in observational studies: the average treatment effect (ATE) and the average treatment effect on the treated (ATT). The ATE measures the average difference in outcomes when all individuals in the study population are assigned to treatment versus when all individuals are assigned to control. On the other hand, the ATT measures the average difference in outcomes among those individuals in the population that were actually exposed to the treatment. The RD estimand, however, is far more local than both of these estimands as it only applies to units close to the cutoff. Ideally, we would like to study more general treatment effects, such as the ATE, in order to learn about the average difference in outcomes that would occur if all units in the study were switched from treated to untreated. Unfortunately, this kind of treatment effects is not generally available in RD designs because the non-experimental treatment assignment only justifies studying effects for units whose scores are near the cutoff---see \cite{Cattaneo-Keele-Titiunik-VazquezBare_2021_JASA} for further discussion and related references. 

\subsubsection{Local Randomization Framework}

The second framework for the analysis of RD designs is based on the idea of local randomization \citep{Cattaneo-Frandsen-Titiunik_2015_JCI,Cattaneo-Titiunik-VazquezBare_2017_JPAM}, where potential outcomes could be viewed as random variables or as fixed quantities, exactly as in the analysis of experiments literature \citep{Rosenbaum_2010_Book,Imbens-Rubin_2015_Book,Hernan-Robins_2022_Book}. To formalize this framework, we introduce notation for the local randomization neighborhood or window, $\W=[\C-w,\C+w]$, where $w>0$ is its half length and $\W$ is assumed symmetric around the cutoff only for simplicity. In this setting, we call $\W$ a \textit{window} to distinguish it from the local neighborhood or bandwidth used in the context of continuity-based methods.

While in the continuity-based framework the key assumption is continuity of conditional expectations to enable extrapolation to the cutoff, in the local randomization framework the idea is to impose conditions to induce an experimental setting near the cutoff. Thus, the key two assumptions are: (i) known treatment assignment mechanism for all units with score in $\W$; and (ii) lack of relationship between score and outcomes for all units with score in $\W$. The second assumption is important. In the continuity-based RD design, the fact that the score is related to the potential outcomes does not present challenges because the parameter of interest is defined at the (single) cutoff point. In contrast, in the local randomization framework, the potential outcomes can be related to the score far from the cutoff, but this relationship must vanish in the window $\W$. As such, we must assume that the value of the score within this interval is unrelated to the potential outcomes---a condition that is not guaranteed by the random assignment of the score $X_i$, nor by the random assignment of the treatment $T_i$. Such an assumption is plausible for small neighborhoods around the cutoff, that is, for those units that have scores closest to the cutoff. See \citet[Section 2]{Cattaneo-Idrobo-Titiunik_2023_Book} and references therein for more discussion and extensions.

In the local randomization framework, we can define treatment effects that are analogous to those discussed in the continuity-based framework. The main difference is that the continuity-based estimands are defined at the cutoff, and the analogous local randomization estimands are defined in the window $\W$ around the cutoff. The local randomization sharp RD parameter is the average treatment effect inside the window $\W$, analogous to $\tau_\mathtt{SRD}$, defined as
\begin{align}
    \theta_\mathtt{SRD} \equiv \E_\W[Y_i(1) - Y_i(0)]
                      = \frac{1}{N_\W} \sum_{i:X_i\in\W} \E_\W\Big[ \frac{T_i Y_i}{\P_\W[T_i=1]} \Big] 
                        - \frac{1}{N_\W} \sum_{i:X_i\in\W} \E_\W\Big[ \frac{(1-T_i) Y_i}{1-\P_\W[T_i=1]} \Big],
\end{align}
where the potential outcomes $Y_i(0)$ and $Y_i(1) $ can be taken as random or fixed depending on the approach taken, $\P_\W[\cdot]$ and $\E_\W[\cdot]$ denote the probability and expectation taken conditionally for those units with $X_i\in \W$, and $N_\W$ is the number of units with $X_i\in \W$. The last expression after the equality sign indicates that $\theta_\mathtt{SRD}$ can be estimated from the data just like in the standard analysis of experiments, but using only units with score within the local randomization window $\W$.

Figure \ref{fig:Local Randomization} showcases the local randomization framework, showing a local neighborhood around $\C$ defined by $\W$. The key idea is that there exists a neighborhood or window around the cutoff where the treatment assignment resembles what it would have been in a randomized experiment. Given this fact, we can simply estimate the treatment effect as if this was an experiment for the units that fall within the local neighborhood around $\C$. As we discuss below, the analogy between RD local randomization and a true experiment is not perfect, and the local randomization RD framework requires stronger assumptions than the continuity-based framework. However, local randomization methods are valid when the score is discrete, while continuity-based methods may be invalid if the score is too coarse (Section \ref{sec:RD Analysis with Discrete Score}).

\subsection{Fuzzy RD Designs}

\begin{figure}[ht]
    \centering
	\begin{subfigure}{0.48\textwidth}
		\centering
		\includegraphics[scale=0.45]{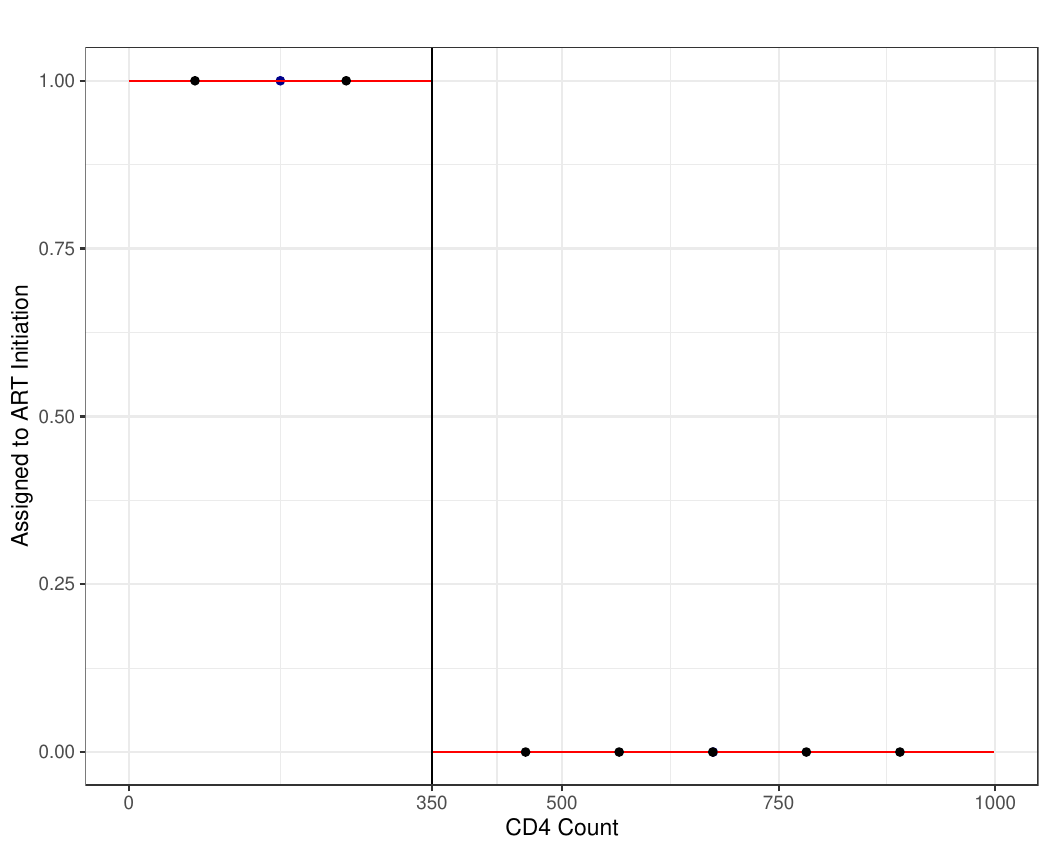}
		\caption{Treatment Assignment ($\mathbb{P}[T_i=1|X_i]$)}\label{fig: ART - Treatment Assignment}
	\end{subfigure}
	\begin{subfigure}{0.48\textwidth}
		\centering
		\includegraphics[scale=0.45]{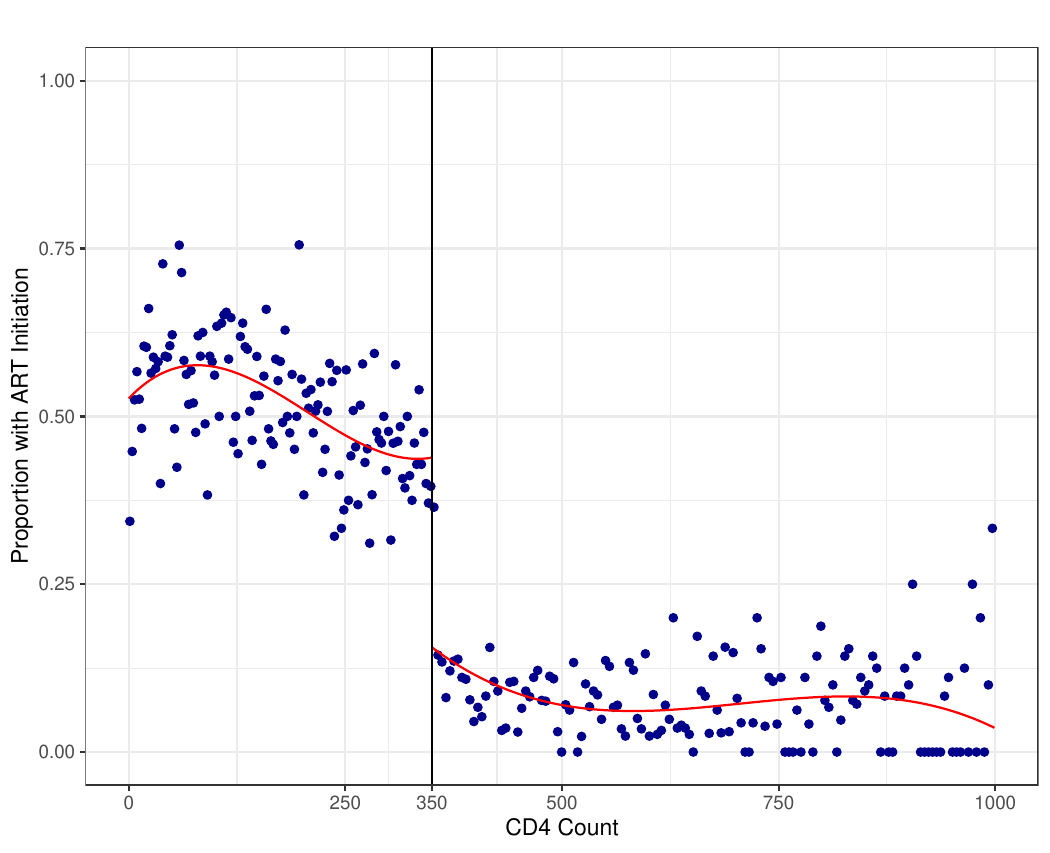}
		\caption{Treatment Take-up ($\mathbb{P}[D_i=1|X_i]$)}\label{fig: ART - Treatment Take-up}		
	\end{subfigure}
	\caption{Treatment Assignment vs. Treatment Take-up --- ART Application}\label{fig: ART - T vs D}
   \flushleft{{\footnotesize Note: The score $X_i$ is patient $i$'s CD4 count, and all patients below the cutoff are assigned to receive ART. In panel (a) the y-axis is $D_i$, an indicator equal to 1 if patient $i$ received ART (0 otherwise); the dots are local means of $D_i$ calculated for patients in different non-overlapping bins of $X_i$, separately above and below the cutoff. In panel (b), the y-axis is $Y_i$, an indicator equal to one if patient $i$ was retained in case (0 otherwise); the dots are local means of $Y_i$ calculated for patients in different non-overlapping bins of $X_i$; and the solid line is a 4th-order polynomial of $Y_i$ on $X_i$, fitted separately for patients above and below the cutoff.}}
\end{figure}

An important feature of the ART application, which is common in many RD designs in biomedical research, is that being \textit{assigned} to the treatment condition is not the same as actually \textit{receiving} the treatment. We use the binary variable $\D_i$ to denote whether the treatment is actually received by unit $i$ ($D_i=1$) or not ($D_i=0$), while we continue to use the binary variable $T_i$ to record whether the treatment is offered ($T_i=1$) or not ($T_i=0$). In the sharp RD design, we always have $T_i=D_i$ because compliance with treatment assignment is perfect, while the defining feature of a fuzzy RD design is that there are some units for which $\T_i\neq \D_i$. For example, in the ART application, there are patients with CD4 counts of less than $350$ ($T_i = \I(X_i < 350)=1$) who never initiate ART ($D_i=0$). This is depicted visually in Figure \ref{fig: ART - T vs D} using RD plots. Figure \ref{fig: ART - Treatment Assignment} shows that all patients with $T_i = \I(X_i < 350)=1$ where assigned to treatment with probability one, while all patients with $T_i = 0$ where assigned to control with probability one. However, treatment assignment was not always followed, as shown in Figure \ref{fig: ART - Treatment Take-up}, which plots the proportion of patients actually receiving ART against the score.

We employ potential outcomes to formalize fuzzy RD designs. Every unit has two \textit{potential treatments}: $D_i(1)$ is the treatment that unit $i$ receives when this unit is assigned to the treatment condition (i.e, when $\T_i =1$), while $D_i(0)$ is the treatment that unit $i$ receives when this unit is assigned to the control condition (i.e, when $\T_i =0$). Both $D_i(1)$ and $D_i(0)$ can be one or zero, depending on unit $i$'s compliance decisions. For example, if unit $i$ is assigned to the treatment condition but refuses to receive the treatment, $D_i(1)=0$; and a unit that complies perfectly with their assignment has $D_i(1)=1$ and $D_i(0)=0$. Thus, the observed treatment received is $D_i = (1-\T_i) \cdot D_i(0) + \T_i \cdot D_i(1)$.

For the outcome of interest, this framework implies that every unit has four different potential outcomes depending on the combination of treatment assignment and compliance decisions: $Y_i(1,0)$, $Y_i(1,1)$, $Y_i(0,0)$, and $Y_i(0,1)$. We denote them generally as $Y_i(T_i, D_i(T_i))$, a function of both the treatment assigned and the treatment received. For example, $Y_i(0,1)$ corresponds to the potential outcome that would occur if unit $i$ were assigned to the control condition ($T_i=0$) but received the treatment anyway ($D_i(0)=1$). However, we only observe the potential outcome and the potential treatment corresponding to the values of $T_i$ and $D_i$ that are realized for unit $i$. Formally, the observed outcome is now $Y_i = (1-\T_i) \cdot Y_i(0, D_i(0)) + \T_i \cdot Y_i(1,D_i(1))$, and the observed data is $(Y_1,D_1,X_1),\ldots,(Y_n,D_n,X_n)$.

\subsubsection{Continuity-Based Framework}

In the fuzzy RD design, the standard RD estimand, $\tau_{\mathtt{SRD}}$, is unavailable except under strong assumptions that will be implausible in many applications (e.g., constant treatment effects as a function of the score). Instead, when there is non-compliance, researchers typically focus on two types of treatment effects: the effects of \textit{assigning} the treatment for all units, and the effect of \textit{receiving} the treatment for a subpopulation of units. Each type of effect requires different assumptions, and which one is of interest depends on the particular application.

The effect of the treatment received is of obvious importance. For example, in the ART application we are interested in the effect of initiating ART on patient retention. However, in some cases, researchers are also interested in the effect of assigning the treatment on the outcome, which is commonly known as the \textit{intention-to-treat} (ITT) effect. This effect includes not only the effect that the treatment received may directly have on the outcome, but also the effect caused by strategic compliance decisions that individuals make in response to knowledge about their assignment. Policy-makers interested in anticipating the overall effects of establishing a new program are often interested in ITT effects.

Within the continuity-based framework, we start by considering the effect of treatment assignment on the outcome ($Y_i$) and on the treatment received ($D_i$), both of which can be seen as sharp RD effects of the treatment assignment. The RD effect of the treatment assignment on the observed outcome is
\begin{equation}
    \tau_{\mathtt{Y}}\equiv\lim_{x\downarrow \C}\E[Y_i|X_i=x]-\lim_{x\uparrow \C}\E[Y_i|X_i=x].
\end{equation}
Under continuity assumptions analogous to those in the canonical sharp RD case, $\tau_{\mathtt{Y}}$ captures the ITT effect of the treatment assignment on the outcome at the cutoff, which we can write as $\tau_{\mathtt{Y}} =\E[Y_i(1,D_i(1))-Y_i(0,D_i(0))|X_i= \C]$, the average change in potential outcomes at the cutoff from switching the assignment from control to treated. In the ART application, this effect is plotted in Figure \ref{fig: ART - RD plot Y-X} using global polynomial approximations. The ITT effect of the treatment assignment on the outcome follows a sharp RD design where the $T_i$ is seen as the treatment of interest. Thus, we estimate the same difference in limits $\lim_{x\downarrow{\C}} \E[Y_i|X_i=x] - \lim_{x\uparrow{\C}} \E[Y_i|X_i=x]$ that we estimate in a sharp RD setting, but we modify the assumptions and interpretation to accommodate imperfect compliance. Because some units fail to comply with their assignment, the sharp RD treatment effect of $T_i$ on $Y_i$ is no longer the effect of the treatment itself, but rather the effect of \textit{assigning} the treatment. For example, in the ART application, $\tau_{\mathtt{Y}}$ captures the average effect of offering ART to patients whose CD4 is $350$ who may or may not accept the offer, while the parameter $\tau_{\mathtt{SRD}}$ would capture the effect of actually starting ART for those patients.

The RD effect of the treatment assignment on the treatment received at the cutoff is
\begin{equation}
    \tau_{\mathtt{D}} \equiv \lim_{x\downarrow \C}\E[D_i|X_i=x]-\lim_{x\uparrow \C}\E[\D_i|X_i=x].
\end{equation}
Since $D_i$ is binary, $\tau_{\mathtt{D}}$  captures the difference in the probability of receiving the treatment at the cutoff between units just assigned to the treatment vs. assigned to the control condition. In the ART application, this is the difference between the proportion of patients with CD4 counts just below 350 who initiate ART and the proportion of patients with CD4 counts just above 350 who initiate ART. This treatment effect is illustrated in Figure \ref{fig: ART - Treatment Take-up}. Under continuity conditions, the difference in treatment probabilities captured by $\tau_{\mathtt{D}}$ can be attributed to the RD assignment rule; in this case, $\tau_{\mathtt{D}}$ represents the average effect of assigning the treatment on receiving the treatment at the cutoff, that is, $\tau_{\mathtt{D}} =  \E[D_i(1)-D_i(0)|X_i=\C]$. This effect is usually called the \textit{first-stage} or \textit{take-up} effect. Both $\tau_{\mathtt{Y}}$ and $\tau_{\mathtt{D}}$ are sharp RD parameters. 

Investigators are often also interested in the effect of receiving the treatment, not merely of assigning it. While $\tau_{\mathtt{SRD}}$ is infeasible in the fuzzy RD design due to non-compliance, under additional assumptions, it is possible to estimate a related parameter that captures the average effect of the treatment at the cutoff for a particular subpopulation of units. We define the fuzzy RD treatment effect as
\begin{equation}
    \tau_{\mathtt{FRD}} \equiv\frac{\tau_{\mathtt{Y}}}{\tau_{\mathtt{D}}},
\end{equation}
which is the ratio of the sharp RD effect of $T_i$ on $Y_i$ and the sharp RD effect of $T_i$ on $D_i$.

The parameter $\tau_{\mathtt{FRD}}$ can be interpreted as the average effect of the treatment received at the cutoff for the subpopulation of units who are compliers---informally defined as units who receive the treatment when their score is above the cutoff and refuse the treatment when their score is below the cutoff. Different authors have formalized the definition of compliers differently in RD settings; a thorough discussion is beyond the scope of our discussion, but we refer the reader to \cite{Dong_2018_OBES}, \cite{Cattaneo-Keele-Titiunik-VazquezBare_2016_JOP}, and \cite{Arai-Hsu-Kitagawa-Mourifie-Wan_2022_QE} for examples, and to \citet[Section 3]{Cattaneo-Idrobo-Titiunik_2023_Book} for a practical discussion. See also \citet{Baiocchi-Cheng-Small_2014_StatsMed} for a review of IV methods for causal inference. 

Regardless of the technical details, interpreting the fuzzy RD parameter as the effect of $D_i$ on $Y_i$ for the compliers requires three assumptions. The formalization of these assumptions varies depending on the particular definitions adopted, but the conceptual ideas are similar in all cases. First, the parameter $\tau_{\mathtt{D}}$ must be nonzero, and well-separated from zero for estimation and inference to be meaningful. In other words, being above versus below the cutoff must induce some units to actually take the treatment. This rules out, for example, a situation where having a CD4 count below 350 induces no patients to start ART. This is usually referred to as the \textit{relevance} assumption or the \textit{first-stage} in IV settings, and is testable. We will showcase this point in Section \ref{sec:Analysis with Continuous Score}.

Second, we need continuity conditions similar to those invoked in the sharp RD case, but generalized for the more complex setting of non-compliance. These continuity conditions will implicitly require, among other things, that the treatment assignment only affect the average outcomes via its effect on the treatment received, but not directly, analogous to the \textit{exclusion} restriction in IV settings. In other words, $T_i$ should only have an affect on $Y_i$ through $D_i$: crossing the cutoff should only affect the outcome if it has an effect of changing the actual treatment received, but not otherwise. This key assumption is untestable and requires careful qualitative reasoning for justification, particularly in medical settings where placebo effects are common \citep{KaptchukMiller2015-NEJM}.

In the ART application, the exclusion restriction requires that having a CD4 count below 350 have no effect on retention in care except by inducing people to initiate ART. This assumption might be implausible if seeing a CD4 count below 350 leads physicians to order additional tests or to communicate with patients differently, which can in turn lead to discovery of other health issues and return for care of a different condition. The exclusion restriction would be more plausible if the outcome were a biological manifestation of HIV rather than retention in care, as it is more plausible that the only way future HIV symptoms would be reduced is through exposure to ART.

Finally, it is also common to assume \textit{monotonicity} or a similar condition for the interpretation of the fuzzy RD treatment effect $\tau_{\mathtt{FRD}}$. Informally, monotonicity requires that a patient who decides to receive the treatment when they are not eligible for it, continues to take the treatment when they are eligible. One way to interpret this in the RD setting where $T_i= \I(X_i\geq \C)$, is to require that a unit with score $X_i$ who refuses the treatment when the cutoff is $\C$ must also refuse the treatment for any cutoff $\C' > \C$, and a unit who takes the treatment when the cutoff is $\C$ must also take the treatment for any cutoff $\C' < \C$. In our example, this implies that a patient who, say, has a CD4 count of $340$ and refuses ART when the cutoff is $350$, he or she must also refuse ART when the cutoff is $330$.

\subsubsection{Local Randomization Framework}

In the local randomization framework for fuzzy RD designs, we can consider parameters of interest that are analogous to those discussed in the continuity-based framework. The sharp RD estimator of the effect of $T_i$ on $Y_i$ and the effect of $T_i$ on $D_i$ are defined, respectively, as
\begin{equation}
    \theta_\mathtt{Y} \equiv \frac{1}{N_\W} \sum_{i:X_i\in\W} \E_\W\Big[ \frac{T_i Y_i}{\P_\W[T_i=1]} \Big] 
                        - \frac{1}{N_\W} \sum_{i:X_i\in\W} \E_\W\Big[ \frac{(1-T_i) Y_i}{1-\P_\W[T_i=1]} \Big]
\end{equation}
and
\begin{equation}
    \theta_\mathtt{D} \equiv \frac{1}{N_\W} \sum_{i:X_i\in\W} \E_\W\Big[ \frac{T_i D_i}{\P_\W[T_i=1]} \Big] 
                        - \frac{1}{N_\W} \sum_{i:X_i\in\W} \E_\W\Big[ \frac{(1-T_i) D_i}{1-\P_\W[T_i=1]} \Big],
\end{equation}
which parallel the continuity-based parameters $\tau_\mathtt{Y}$ and $\tau_\mathtt{D}$. Under the local randomization assumptions, the parameters $\theta_\mathtt{Y} $ and $\theta_\mathtt{D}$ capture the average effect of assigning the treatment for observations with scores in the window. Finally, we can also define the local-randomization fuzzy RD parameter as the ratio: $\theta_\mathtt{FRD} \equiv \theta_\mathtt{Y} /\theta_\mathtt{D}$.

As in the continuity-based framework, under appropriate assumptions, $\theta_\mathtt{FRD}$ can be interpreted as the average treatment effect in the window for compliers. The assumptions typically used are similar to those required in IV settings, now applied to observations with scores in the window $\W$, and hence similar to those discussed for the continuity-based framework. Once again, the effect of the treatment assignment on the treatment received, $\theta_\mathtt{D}$, must be well separated from zero. The exclusion restriction that the treatment assignment have no direct effect on the outcomes must also hold for all units with scores within the window; this restriction is implied by the local randomization condition that the (distribution of the) potential outcomes and potential treatments is not a function of the score inside $\W$. Finally, the assumption of monotonicity requires that there be no units with scores in $\W$ who receive a treatment condition that is always opposite to their assignment.

Finally, it is important to understand how to interpret the fuzzy RD estimands $\tau_\mathtt{FRD}$ and $\theta_\mathtt{FRD}$. These estimands differ from the (local) average treatment effects that are commonly used in the IV literature \citep{Baiocchi-Cheng-Small_2014_StatsMed}: the fuzzy RD estimands capture the average treatment effect for a subpopulation (e.g., compliers) with a score value at or near the cutoff, and by implication often have lower external validity than the standard IV estimand. In the ART application, the fuzzy RD treatment effects only apply to the set of compliers with scores near $350$. The fuzzy RD treatment effect may differ compared to those patients with much higher or lower CD4 counts. For more discussion on extrapolation of RD treatment effects away from the cutoff, see \citet{Cattaneo-Keele-Titiunik-VazquezBare_2021_JASA} and references therein.

\section{Analysis with Continuous Score}\label{sec:Analysis with Continuous Score}

We now discuss estimation, inference, and validation methods within the continuity-based and the local randomization RD frameworks with a continuously distributed score, using again the ART application as the running empirical example. All the results in this section can be reproduced using the replication materials.

\subsection{Continuity-Based Methods}

A common problem in RD settings is that there are often few observations with score values very close to the cutoff, which means that estimating the effect at $X_i=\C$ requires using observations whose values of $X_i$ are relatively far from $\C$. Because a sufficiently smooth function can be well approximated by a polynomial function, up to misspecification error, standard continuity-based RD estimation methods approximate the regression functions, $\E[Y_i(0) | X_i = x]$ and $\E[Y_i(1) | X_i = x]$ using a polynomial function of the score. 

\subsubsection{Point Estimation}

Modern RD estimation is based on local polynomial approximations that discard observations sufficiently far away from the cutoff and then employ a low-order polynomial approximation (usually linear or quadratic) for estimation. This approach is known as local polynomial regression in the statistical literature \citep{Fan-Gijbels_1996_Book}. State-of-the-art RD methods use two separate linear polynomial fits for treated and control units using only observations near the cutoff as determined by the choice of a bandwidth parameter. This local approach is more robust and less sensitive to boundary and over-fitting problems.

Local polynomial methods require the user to make three choices: the bandwidth, the kernel function, and the polynomial order. The bandwidth controls the width of the neighborhood around the cutoff that is used to fit the local polynomial models, and hence determines the number of observations above and below the cutoff that are used for estimation. Within the neighborhood determined by the bandwidth, it is common to adopt a weighting scheme to ensure that the observations closer to $\C$ receive more weight than those further away. The weighting scheme is referred to as a kernel function, $K(\cdot)$, and two common options are the triangular kernel, $K(x)=(1-|x|)\I(|x|\leq 1)$, which linearly down-weights observations within the bandwidth, and the uniform kernel, $K(x)=\I(|x|\leq 1)$, which gives equal weight to all observations within the bandwidth. The polynomial order $p$ determines the order of the polynomial approximation near the cutoff. In the software resources used in this tutorial, the defaults are linear fit ($p=1$) and triangular kernel. These choices have objective theoretical advantages in the nonparametrics literature \citep{Fan-Gijbels_1996_Book}, but the researcher can also investigate the robustness of the empirical results by choosing $p=2$ or a uniform kernel.

The RD estimate is thus constructed as follows. For observations above the cutoff (i.e., observations with $X_i \geq \C$), fit a weighted least squares regression of the outcome $Y_i$ on a constant and $(X_i-\C), (X_i-\C)^2,\ldots, (X_i-\C)^p$ with weight $K(\frac{X_i-\C}{h})$ for each observation, leading to the estimated equation $\widehat{Y}_i = \widehat{\mu}_+ + \widehat\mu_{+,1} (X_i-\C) + \widehat\mu_{+,2} (X_i-\C)^2 + \cdots + \widehat \mu_{+,p} (X_i-\C)^p$, where the estimated intercept, $\widehat{\mu}_+$, is a point estimate of $\mu_+=\E[Y_i(1)|X_i=\C]$. Similarly, for observations below the cutoff, fit a weighted least squares regression of the outcome $Y_i$ on a constant and $(X_i-\C), (X_i-\C)^2,\ldots, (X_i-\C)^p$ with weight $K(\frac{X_i-\C}{h})$ for each observation, leading to $\widehat{Y}_i = \widehat{\mu}_- + \widehat\mu_{-,1} (X_i-\C) + \widehat\mu_{-,2} (X_i-\C)^2 + \cdots + \widehat \mu_{-,p} (X_i-\C)^p$, where the estimated intercept, $\widehat{\mu}_-$, is a point estimate of $\mu_-=\E[Y_i(0)|X_i=\C]$. Therefore, the sharp RD point estimate is
\begin{equation}
    \widehat{\tau}_\mathtt{SRD} = \widehat{\mu}_+ - \widehat{\mu}_-.
\end{equation}

The choice of bandwidth $h$, which determines which observations near the cutoff are used, is the most critical  when implementing local polynomial RD methods. A small bandwidth will reduce the approximation error of the local polynomial approximation because it only uses observations very close to the cutoff. However, a small bandwidth will also increase the variance of the estimates because only a few observations are used in the local fit. Analogously, a large bandwidth may increase the approximation error if the underlying regression function differs considerably from the polynomial approximation used, but will result in lower variance due to the relatively larger number of observations included. Thus, bandwidth selection embodies a bias-variance trade-off: smaller bandwidths will tend to have less bias but higher variance, and viceversa. The mean squared error (MSE) of any estimator is the sum of its bias squared plus its variance; given the bias-variance tradeoff, bandwidth selection can be automated in a principled, data-driven way by first deriving an approximation to the MSE of the RD point estimator, and then choosing the value of $h$ that minimizes it. This so-called MSE-optimal bandwidth selection approach has become the standard for RD estimates. See \citet{Calonico-Cattaneo-Titiunik_2014_ECMA,Calonico-Cattaneo-Farrell-Titiunik_2019_RESTAT} for the most recent methodological developments, and \cite{Cattaneo-VazquezBare_2016_ObsStud} and \cite{Calonico-Cattaneo-Farrell_2020_ECTJ} for an overview on neighborhood selection methods in RD designs more generally.

\subsubsection{Confidence Intervals}

The MSE-optimal bandwidth is used to construct an MSE-optimal point estimator, $\widehat{\tau}_\mathtt{SRD}$, but using that bandwidth to conduct standard least squares inference is in general \textit{invalid} \citep{Calonico-Cattaneo-Titiunik_2014_ECMA,Calonico-Cattaneo-Farrell-Titiunik_2019_RESTAT,Calonico-Cattaneo-Farrell_2020_ECTJ}. To be more precise, the MSE-optimal bandwidth balances bias and variance in such a way that the point RD estimator exhibits a misspecification bias in its distribution, which leads to confidence intervals and hypothesis tests that are invalid in general, even in large samples. This implies that the usual asymptotic $95$-percent confidence interval for $\tau_\mathtt{SRD}$ given by $\mathtt{CI} = \big[\widehat{\tau}_\mathtt{SRD} \pm 1.96 \cdot \sqrt{\widehat{\mathtt{V}}} \big]$, where $\widehat{\mathtt{V}}$ denotes a variance estimator, is invalid because the underlying Gaussian distribution of the RD point estimator has a non-zero bias when the MSE-optimal bandwidth is used. It can be shown that $\mathtt{CI}$ will cover the population treatment effect $\tau_\mathtt{SRD}$ roughly $80$\% of the time in repeated sampling, implying a false rejection rate of about $15$-percentage points.

A principled alternative is to use the \textit{robust bias corrected} confidence intervals proposed by \citet{Calonico-Cattaneo-Titiunik_2014_ECMA}, and later extended to other settings \citep{Xu_2017_JoE,Arai-Ichimura_2018_QE,Calonico-Cattaneo-Farrell-Titiunik_2019_RESTAT,Dong-Lee-Gou_2021_JASA,Arai-Otsu-Seo_2021_wp}. Robust bias corrected confidence intervals modify the classical confidence intervals $\mathtt{CI}$ in two ways: (i) the point estimator $\widehat{\tau}_\mathtt{SRD}$ is debiased by including an estimate of the leading misspecification error (denoted by $\widehat{\mathtt{B}}$), and (ii) the variance estimator $\widehat{\mathtt{V}}$ is increased to incorporate the contribution of the bias correction step to the overall variability of the confidence interval (denoted by $\widehat{\mathtt{W}}$). Thus, the robust bias corrected confidence intervals take the form
\begin{equation*}
 \mathtt{CI}_{\mathtt{RBC}}=\left[ ~\left( \widehat{\tau}_\mathtt{SRD}-\widehat{\mathtt{B}}\right)\pm1.96\cdot \sqrt{\widehat{\mathtt{V}}+\widehat{\mathtt{W}}}~\right].   
\end{equation*}

These confidence intervals are valid even when the MSE-optimal bandwidth is used, and have several demonstrable theoretical properties, including smaller coverage errors and less sensitivity to tuning parameter choices \citep{Calonico-Cattaneo-Farrell_2018_JASA,Calonico-Cattaneo-Farrell_2022_Bernoulli,Kamat_2018_ET,Tuvaandorj_2020_JoE}. Furthermore, the improved finite sample performance of these intervals has been validated empirically \citep{Ganong-Jager_2018_JASA,Hyytinen-etal_2018_QE,DeMagalhaes-etal_2020_wp}.

Our practical recommendation is therefore to (i) report the MSE-optimal RD point estimate $\widehat{\tau}_\mathtt{SRD}$, which is constructed using an MSE-optimal bandwidth choice, and (ii) report robust bias corrected confidence intervals, which employ the same MSE-optimal bandwidth choice. All these methods are readily available in \texttt{Python}, \texttt{R}, and \texttt{Stata} general-purpose software packages (\url{https://rdpackages.github.io/}). We use these methods for the analysis of our three empirical examples, as illustrated in the accompanying replication files. 

The local polynomial methods for sharp RD continuity-based analysis can be extended to fuzzy RD designs to estimate $\tau_{\mathtt{Y}}$, $\tau_{\mathtt{D}}$, and $\tau_{\mathtt{FRD}}$. The first point estimator is exactly the same as described above, using local polynomials to estimate the relationship between $Y_i$ and the score $X_i$---that is, $\widehat{\tau}_{\mathtt{Y}}=\widehat{\tau}_{\mathtt{SRD}}$. The estimator $\widehat{\tau}_{\mathtt{D}}$ of $\tau_{\mathtt{D}}$ is constructed analogously, after replacing the observed outcome variable $Y_i$ with the observed treatment status $D_i$. Once  $\widehat{\tau}_{\mathtt{Y}}$ and $\widehat{\tau}_{\mathtt{D}}$ are available, the fuzzy RD estimand $\tau_{\mathtt{FRD}}$ is estimated using $\widehat{\tau}_{\mathtt{FRD}} = \widehat{\tau}_{\mathtt{Y}}/\widehat{\tau}_{\mathtt{D}}$.

The estimator $\widehat{\tau}_{\mathtt{FRD}}$ is consistent for $\tau_{\mathtt{FRD}}$ under standard regularity conditions, although it may exhibit more bias or other potential problems due to its intrinsic ratio structure. Heuristically, everything discussed in this section still applies to this estimator, but some more details are necessary. First, bandwidth selection can still proceed based on a MSE approximation, although now such approximation should also take into account the ratio structure of the estimator. Furthermore, more than one natural MSE-optimal bandwidth choice is available: it is possible to consider one single bandwidth for the ratio $\widehat{\tau}_{\mathtt{FRD}}$, or two distinct bandwidth choices, one each for the numerator and denominator. In practice, most researchers employ a single MSE-optimal choice for $\widehat{\tau}_{\mathtt{FRD}}$ or for $\widehat{\tau}_{\mathtt{Y}}=\widehat{\tau}_{\mathtt{SRD}}$, although some researchers prefer to choose two different bandwidths for $\widehat{\tau}_{\mathtt{Y}}$ and $\widehat{\tau}_{\mathtt{D}}$. As a general rule, it is usually recommended to use a single MSE-optimal bandwidth for the estimator of interest, in this case, $\widehat{\tau}_{\mathtt{FRD}}$. For inference, the same problems of misspecification biases arise in the fuzzy RD design, usually made more acute by the ratio structure of the point estimator. As a consequence, robust bias correction continues to be recommended whenever an MSE-optimal bandwidth choice is used for point estimation. Because these formulas are cumbersome we do not reproduce them here, but they can all be found in \cite{Calonico-Cattaneo-Titiunik_2014_ECMA,Calonico-Cattaneo-Farrell-Titiunik_2019_RESTAT,Calonico-Cattaneo-Farrell_2020_ECTJ}.

\subsubsection{Continuity-Based Analysis of ART Example}

We now illustrate all the methods discussed so far using the ART application. The effects on the main outcome of interest are reported in Table~\ref{tab:rdd.hiv1}. All the results in this table can be generated using \texttt{rdrobust} in any of the three software platforms (\texttt{Python}, \texttt{R}, \texttt{Stata}). First, we focus on the effect of being assigned to treatment (in this case, having a score below the cutoff). We find that having a CD4 count of $350$ or greater reduces the likelihood of ART initiation by $21$ percentage points ($\widehat{\tau}_{\mathtt{D}}$) and also reduces program retention by $14$ percentage points ($\widehat{\tau}_{\mathtt{Y}}$). This means that being just below the $350$ threshold increases likelihood of both ART and program retention. These are the effects of assignment to ART rather than of actual ART initiation, and as such do not fully capture the primary effect of interest---the effect of ART initiation on program retention. To explore the latter effect, we focus on the fuzzy RD estimate, $\widehat{\tau}_{\mathtt{FRD}}$, which is simply the ratio of the two ITT effects. We find that ART initiation increases program retention by more than 67 percentage points, and the confidence interval is bounded away from zero. Thus, we conclude that patients who initiated ART were much more likely to be retained in the treatment program. Under standard fuzzy RD assumptions, this is the effect on program retention of initiating ART for patients with a CD4 count of 350 who are compliers. Recall that this interpretation requires, among other assumptions, that there is no effect of having a CD4 count below 350 on patient retention except via ART initiation.

We note that the analysis of RD designs can be enhanced by including predetermined covariates, which can be incorporated in a variety of ways. As in randomized experiments, a natural use of covariates is to improve the efficiency of the local polynomial RD estimator, as developed in \citet{Calonico-Cattaneo-Farrell-Titiunik_2019_RESTAT}. However, predetermined covariates cannot be used to salvage an invalid RD design because incorporating covariates on those settings necessarily changes the RD parameters interest. See \citet{Cattaneo-Keele-Titiunik_2023_HandbookCh} for more discussion and references.

\begin{table}[ht]
\begin{center}
\resizebox{\textwidth}{!}{
    \begin{threeparttable}
    \caption{RD Estimation and Inference --- ART Application}\label{tab:rdd.hiv1}
    \begin{tabular}{lccccc}
        \toprule
        \multicolumn{6}{c}{Continuity-Based Methods} \\
        \midrule
         & RD Effect & 95\% Robust CI & Bandwidth ($h$) &  $N^-_h$ & $N^+_h$\\ 
          \midrule
          ITT Effect of ART Assignment on ART Initiation & -0.21 & [-0.28,-0.12] & 114.36 & 1494 & 1188 \\ 
          ITT Effect of ART Assignment on Program Retention & -0.14 & [-0.22,-0.05] & 114.36 & 1494 & 1188 \\ 
          Fuzzy Effect of ART Initiation on Program Retention & 0.67 & [0.34,1] & 114.36 & 1494 & 1188 \\ 
        \midrule  
        \multicolumn{6}{c}{Local Randomization Methods} \\
        \midrule  
          & Risk Difference & 95\% Confidence Interval & Window ($\W$) & $N_\W^-$ &  $N_\W^+$ \\ 
          \midrule
          ITT Effect of ART Assignment on ART Initiation & 0.02 & [-0.15 , 0.19] & [346,354] & 62 & 58 \\ 
          ITT Effect of ART Assignment on Program Retention & -0.02 & [-0.2 , 0.16] & [346,354] & 62 & 58 \\ 
          Fuzzy Effect of ART Initiation on Program Retention & -0.8 & [-12.5 , 10.91] & [346,354] & 62 & 58 \\  
          \bottomrule
    \end{tabular}
    \begin{tablenotes}[para]
        Note: The first three rows show, respectively, $\widehat{\tau}_{\mathtt{D}}$, $\widehat{\tau}_{\mathtt{Y}}$, and $\widehat{\tau}_{\mathtt{FRD}}$, corresponding to the continuity-based estimates based on local linear estimation with MSE-optimal main bandwidth reported in third column. Column labeled ``95\% Robust CI'' reports the robust 95\% confidence intervals based on robust bias-corrected inference. Column $N^-_h$ reports the number of observations with score in $[\C-h,\C)$ and column $N^+_h$ reports the number of observations with score in $[\C,\C+h]$. The last three show, respectively, $\widehat{\theta}_{\mathtt{D}}$, $\widehat{\theta}_{\mathtt{Y}}$, and $\widehat{\theta}_{\mathtt{FRD}}$, corresponding to the local randomization estimates based on the data-driven chosen local randomization window reported in third column. Column $N^-_\W$ reports the number of observations with score in $\W$ and below the cutoff ($T_i=0$) and column $N^+_\W$ reports the number of observations with score in $\W$ and above the cutoff ($T_i=1$).
    \end{tablenotes}
    \end{threeparttable}
}
\end{center}
\end{table}

\subsection{Local Randomization Methods}

The practical implementation of the local randomization framework requires two steps: (i) choosing the window $\W$ where the local randomization conditions are assumed to hold, and (ii) deploying methods from the analysis of experiments to perform estimation and inference for observations whose scores are inside the window.

Window selection is the most important step in the implementation of the local randomization approach for RD analysis. Although $\W$ could be selected in an ad-hoc fashion, a more principled approach is to select it using predetermined covariates, as proposed by \cite{Cattaneo-Frandsen-Titiunik_2015_JCI}. See also \citet{Cattaneo-Titiunik-VazquezBare_2017_JPAM} and \citet[Section 2]{Cattaneo-Idrobo-Titiunik_2023_Book}.

This data-driven window selection method requires that there be a set of predetermined covariates, $Z$, that are related to the score everywhere except inside $\W$. Once these predetermined covariates are chosen, the implementation of the window selection can be based on methods that assume random sampling (usually called super-population methods) or methods that condition on the units in the sample and assume that the only randomness comes from the treatment assignment mechanism (called Fisherian methods after statistician Ronald Fisher). For an in-depth review of super-population versus Fisherian methods in the causal inference framework, see \cite{Rosenbaum_2010_Book,Imbens-Rubin_2015_Book}. Unlike super-population methods, Fisherian inference methods are exact in finite samples. Thus, in the context of RD window selection, it is often preferable to employ Fisherian methods because the windows considered typically have very few observations, which can invalidate the use of large-sample approximations. 

For implementation, the researcher chooses a test statistic and performs a sequence of hypothesis tests that test the null hypothesis that the treatment has no effect on the covariates inside the window. The first test is conducted in the smallest window around the cutoff that has enough observations (typically a minimum of at least 10 observations on either side is recommended); the sequence continues testing the null hypothesis of no treatment effect on $Z$ in progressively larger windows until this hypothesis is rejected. While clearly this methodology relies on multiple hypothesis testing, there is no need to adjust the inferences because  over-rejection of the null hypothesis leads to a more conservative window choice (i.e., a smaller one). Consequently, a recommended rule is to reject all windows leading to p-values smaller than $0.15$ or $0.10$---these recommendations are based on power calculations under specific assumptions. (When  $Z$ includes multiple covariates, researchers can use the single p-value from an omnibus balance test, or the minimum p-value across individual balance tests.) The chosen $\W$ is the largest (symmetric) interval around the cutoff such that the predetermined covariates of the units inside the window are balanced between treated and control in that window, and in all smaller windows contained in it.

Once window selection is complete, analysis within the local randomization framework is straightforward. Under super-population methods, for example, a natural estimator for $\theta_\mathtt{SRD}$ is the difference in means between the observed outcomes in the treated and control groups. When compliance is imperfect, we can estimate the sharp RD effects of $T_i$ on $Y_i$ and $D_i$, $\theta_\mathtt{Y}$ and $\theta_\mathtt{D}$, with the difference in the average observed outcomes between the treated and control groups inside the window, denoted by $\widehat{\theta}_\mathtt{Y}$ and $\widehat{\theta}_\mathtt{D}$, respectively. We can then estimate the local randomization fuzzy RD parameter, $\theta_\mathtt{FRD}$, with $\widehat{\theta}_\mathtt{FRD} = \widehat{\theta}_\mathtt{Y}/\widehat{\theta}_\mathtt{D}$. In the super-population framework, statistical inferences are based on standard large-sample approximations. In the specific context of RD, this means that the number of units within $\W$ is assumed to be large enough for distributional approximations to hold. This approach directly justifies the use of confidence intervals and p-values based on the large-sample properties of common test statistics such as standardized difference-in-means, least-squares and two-stage least-squares coefficients, etc., frequently used in the analysis of experiments.

When the number of observations in $\W$ is small, adopting a Fisherian approach is more appropriate. This approach takes potential outcomes as non-random and assumes that the randomization mechanism that assigned units to treated and control is either known or can be approximated. The fixed-margins assignment assumption is a natural choice. Fisherian randomization inference employs the sharp null hypothesis of no treatment effect for any unit within $\W$, controlling Type I error for any sample size. Most applications employ the difference-in-means between control and treatment units as the test statistics, but other choices are possible. Under additional assumptions on the treatment effect structure, point estimators and confidence intervals can also be constructed. For example, if we assume $Y_i(1) = Y_i(0) + \tau$, called a constant treatment effect model, we can form a point estimator and confidence intervals for $\tau$ based on standard Fisherian methods. Fisherian methods are also available for fuzzy RD designs where compliance is imperfect. See \cite{Ernst_2004_SS} for a review on permutation-based methods, \cite{Cattaneo-Frandsen-Titiunik_2015_JCI,Cattaneo-Titiunik-VazquezBare_2017_JPAM,Cattaneo-Idrobo-Titiunik_2023_Book} for more details on local randomization RD analysis, and \cite{Keele-Small-Grieve_2017_JRSSA} and \cite{Kang-Peck-Keele_2018_JRSSA} for related methodological developments for IV designs, which could be developed in the context of RD designs.

We illustrate the local randomization RD approach with the ART application. The analysis begins with window selection. The \texttt{rdlocrand} package contains tailored functions for the analysis of RD designs under local randomization, including a function for window selection. Using the data-driven methods for window selection outlined above, the window selected is $[346, 354]$---that is, we find that predetermined covariates in the data are balanced for patients with CD4 counts between $346$ and $354$. We omit the balance test results for space considerations, but the window selection was based on the same covariates shown in Table \ref{tab:cd1}. The resulting local randomization window is much narrower than the neighborhood implied by the bandwidth estimated using continuity-based methods, which is a common phenomenon in practice. The local randomization approach leads to a local neighborhood $\W=[346,354]$ with $121$ patients, while the estimated bandwidth from the continuity-based analysis in Table \ref{tab:rdd.hiv1} leads to the local region $[239,461]$ with $2,593$ patients. By their very nature, local randomization methods focus on much smaller neighborhoods around the cutoff, and thus use substantially fewer observations, which implies that they generally have less statistical precision than continuity-based methods. Nevertheless, local randomization methods can offer a useful complement and a robustness check for continuity-based methods when both frameworks are applicable. 

Table \ref{tab:rdd.hiv1} presents the main empirical results for the ART application using local randomization methods. Given the small sample sizes, all estimated effects are statistically indistinguishable from zero at conventional levels. But the confidence intervals do cover the point estimates reported with the continuity-based methods in Table \ref{tab:rdd.hiv1}. Thus, the results are statistically consistent with each other, albeit the local randomization methods are less informative than the continuity-based methods. One way to further investigate the role of lack of statistical precision is to increase the local randomization neighborhood. We regard this approach as a sensitivity test for the RD design, so we discuss it further below along with other falsification methods. Table \ref{tab:lr--W} reports results for three wider windows---$[340,360]$, $[335,365]$, and $[330,370]$; the results are already closer to results obtained with continuity-based methods in the smallest window, and nearly identical in the other two windows.

\subsection{Evaluating the RD Assumptions}

While one the strongest methods for causal inference and program evaluation, RD designs are ultimately a type of observational study and their key underlying assumptions are not guaranteed to hold by design \citep{Sekhon-Titiunik_2016_ObsStud,Sekhon-Titiunik_2017_AIE}. The main threat to the validity of any RD design is the possibility of the units changing or ``manipulating'' their score in order to systematically select into the treatment. Analysts can offer supporting evidence in favor of the validity of the RD design in two main ways. First, investigators should provide qualitative information about the administrative process by which scores are assigned and cutoffs are determined---including whether this information is public knowledge. In medical applications, we might expect the RD design to be more robust when the score is a lab test. For example, in the ART example, patients might try to influence their CD4 count in order to qualify for ART. However, so long as the CD4 count is determined by laboratory procedures that cannot be precisely manipulated by patients or physicians, this is not a concern.

Second, the analysis of RD designs should include a series of falsification tests and diagnostics. As a general rule, falsification tests cannot prove that an assumption holds, but they can provide indirect empirical evidence that an assumption is likely to be invalid. Falsification tests arise from the fact that causal theories often predict an absence of treatment effects in addition to predicting the presence of such effects. We review several key falsification and diagnostic tests for RD designs, and illustrate their use with the ART application. All these methods are applicable to both sharp and fuzzy RD settings. In addition, similarly to IV settings, we stress the importance of checking the strength of the first-stage estimate in the fuzzy RD design (i.e., $\tau_\mathtt{D}$ and $\theta_\mathtt{D}$ should be well-separated from zero). See \citet{Cattaneo-Idrobo-Titiunik_2020_Book,Cattaneo-Idrobo-Titiunik_2023_Book} for more discussion, and \citet{Cattaneo-Titiunik_2022_ARE} for an overview of the literature.

\subsubsection{Score Density near the Cutoff}

This diagnostic test examines whether, in a local neighborhood near the cutoff, the number of observations below the cutoff is surprisingly different from the number of observations above it \citep{McCrary_2008_JoE}. The underlying assumption is that if individuals do not have the ability to precisely manipulate the value of the score that they receive, the number of treated observations just above the cutoff should be approximately similar to the number of control observations below it. Although this assumption is neither necessary nor sufficient for the validity of an RD design, RD applications where there is an unexplained abrupt change in the number of observations at the cutoff will tend to be less credible.
    
This test is usually implemented in two ways, each motivated by one of the main two RD frameworks discussed in previous sections. The first method is the \textit{Binomial Test} introduced by \cite{Cattaneo-Frandsen-Titiunik_2015_JCI,Cattaneo-Titiunik-VazquezBare_2017_JPAM}, building on the local randomization framework. The second method is the \textit{Density Test} introduced by \cite{McCrary_2008_JoE}, which is based on the continuity-based framework for RD analysis. Informally, both tests seek to detect whether there is a significant amount of ``bunching'' at or near the cutoff. A small p-value under both tests indicates a significant amount of bunching, which is a concern. \cite{Cattaneo-Jansson-Ma_2020_JASA} develop a version of the density test based on local polynomial density estimation which can be plotted. All these tests are available in the software resources.

\begin{figure}[h]
     \centering
     \begin{subfigure}[b]{0.48\textwidth}
        \centering
        \centerline{\includegraphics[scale=0.45]{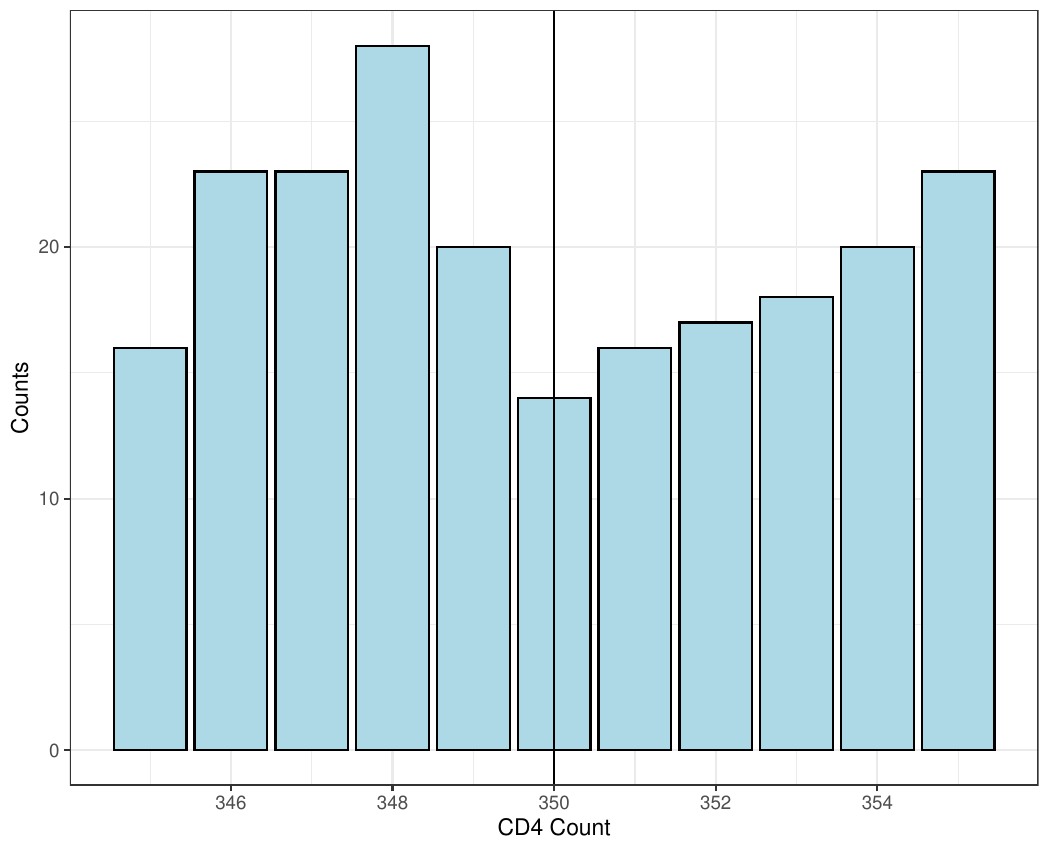}}
        \caption{Histogram near cutoff: Score ($X_i$) within window $[345,355]$.}
        \label{fig.hiv5}
     \end{subfigure}\hfill
     \begin{subfigure}[b]{0.48\textwidth}
        \centering
        \centerline{\includegraphics[scale=0.45]{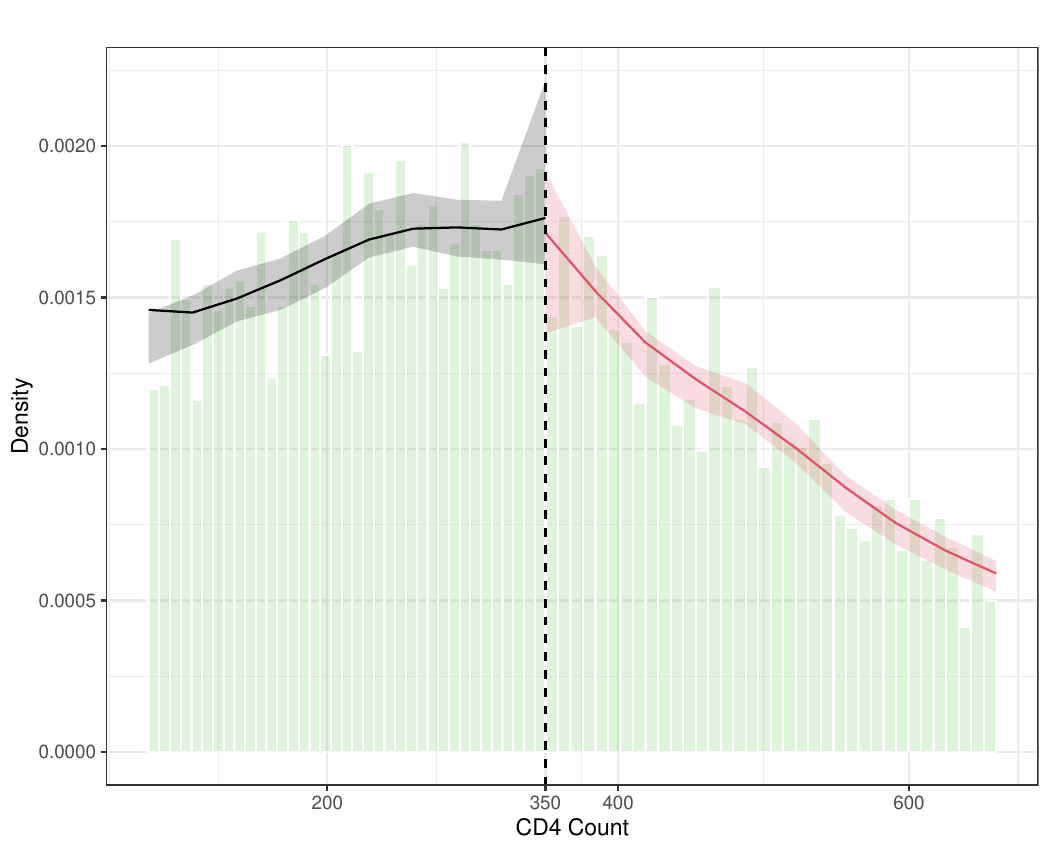}} 
        \caption{Histogram and local polynomial density estimate for score ($X_i$) with $95\%$ robust bias corrected confidence intervals.}
        \label{fig.hiv.dens}
     \end{subfigure}
     \caption{Density Plots for CD4 count (score) around the cutoff --- ART application.}
     \label{fig:Density-HIV}
        \flushleft{{\footnotesize Note: The score $X_i$ is patient $i$'s CD4 count, and all patients below the cutoff are assigned to receive ART. In panel (b), the solid line is a local polynomial estimate of the density of $X_i$ and the shaded regions represent 95\% confidence intervals, both calculated separately for patients above and below the cutoff.}}
\end{figure}

Figure \ref{fig: ART - Histogram X} showed the raw histogram of the score in the ART application; in Figure \ref{fig.hiv5}, we zoom in, showing the histogram only for the region $[345,355]$. Informally, there appear to be no obvious signs of bunching near the $350$ cutoff. Formally, we do not reject the hypothesis of a change in density near the cutoff using the binomial test in the window $[349,350]$ (p-value $=0.2026$). We also implement the density test based on local polynomials, which we illustrate in Figure \ref{fig.hiv.dens}. We fail to reject the null hypothesis that, at the cutoff, the limit of the score density from above the cutoff is the same as the limit from below ($p = 0.1858$). (For implementation, we used the \texttt{rddensity} and the \texttt{rdlocrand} software packages.) Overall, we find no evidence that the density of the score changes abruptly at or near the $350$ cutoff and thus we see no evidence of intentional manipulation of the score.

\subsubsection{Predetermined Covariates and Placebo Outcomes}

Another important falsification test is based on the idea that if units lack the ability to precisely manipulate the value of their score, units just above and just below the cutoff should be similar in terms of all characteristics that could not have been affected by the treatment. These characteristics can be divided into two groups: \textit{predetermined covariates}---variables that are determined before the treatment is assigned, and \textit{placebo outcomes}---variables that are determined after the treatment is assigned but, according to substantive knowledge, could not have been affected by the treatment. In general, baseline covariates should be available in most applications, but the availability of placebo outcomes will vary from application to application.

This falsification test consists of repeating the RD analysis with baseline covariates or placebo outcomes in place of the main outcome of interest. The implementation can be done using both the continuity-based and local randomization frameworks. The underlying assumptions and methods for each case are analogous to those described previously, with the only change that now the outcome variable is either a predetermined covariate or a placebo outcome. As such, implementing this falsification test does not require any special software resources other than those for standard RD estimation and inference. With continuity-based methods, the implementation should use a bandwidth that is specific to each baseline covariate or placebo outcome, instead of the bandwidth selected for the main outcome of interest. In the local randomization framework, some predetermined covariates $Z$ are used to select the window while others could be used for falsification testing after the local randomization window is selected; all falsification tests are conducted in the same chosen window. Regardless of the specific framework and methods employed, from the perspective of falsifying the RD design using predetermined covariates or placebo outcomes, the null hypothesis of no treatment effect should not be rejected in order to offer empirical evidence in favor of the RD assumptions.

We illustrate these ideas with the ART application, estimating the RD treatment effect for predetermined covariates. The results are based on the function \texttt{rdrobust} in the \texttt{rdrobust} package, which is used for estimation of RD effects and includes both bandwidth selection and robust bias correction inference methods. Table \ref{tab:cd1} analyzes the available baseline covariates in the data. The results are calculated using robust local polynomial methods to estimate RD effects, treating each predetermined covariate as an outcome. We find that covariate differences at the cutoff are generally quite small and none of the p-values are below $0.10$. These results are reassuring, as they do not show signs of systematic differences near the cutoff: patients just above the cutoff are similar in terms of baseline covariates to patients just below the cutoff. Similar results are obtained when using local randomization methods, which we omit to conserve space.

\begin{table}[ht]
\begin{center}
\resizebox{\textwidth}{!}{
    \begin{threeparttable}
    \caption{Continuity-Based ITT RD Estimates for Predetermined Covariates with Robust Bias Corrected Inference --- ART Application}\label{tab:cd1}
    \begin{tabular}{lccccccc}
        \toprule
         & Mean Below & Mean Above & $\widehat{\tau}_{\mathtt{Y}}$ & Robust p-value & MSE-Optimal Bandwidth &  $N^-_h$ & $N^+_h$ \\ 
          \toprule
        Age 0-18 & 0.07 & 0.08 & 0.01 & 0.44 & 126.62 & 2389 & 1893 \\ 
          Age 18 -25 & 0.27 & 0.30 & 0.03 & 0.53 & 116.23 & 2178 & 1759 \\ 
          Age 25-30 & 0.24 & 0.19 & -0.04 & 0.16 & 153.19 & 2860 & 2223 \\ 
          Age 30-35 & 0.14 & 0.14 & -0.01 & 0.84 & 109.77 & 2056 & 1653 \\ 
          Age 35-40 & 0.09 & 0.10 & 0.01 & 0.69 & 143.56 & 2689 & 2102 \\ 
          Age 40-45 & 0.07 & 0.07 & -0.00 & 0.93 & 106.65 & 1992 & 1617 \\ 
          Age 45-55 & 0.10 & 0.09 & -0.00 & 0.81 & 131.88 & 2463 & 1953 \\ 
          Age 55+ & 0.04 & 0.03 & -0.01 & 0.50 & 92.74 & 1733 & 1457 \\ 
          2011 Qtr3 & 0.13 & 0.11 & -0.03 & 0.23 & 139.49 & 2666 & 2092 \\ 
          2011 Qtr4 & 0.18 & 0.17 & -0.01 & 0.68 & 108.15 & 2085 & 1668 \\ 
          2012 Qtr1 & 0.19 & 0.21 & 0.02 & 0.49 & 158.60 & 2982 & 2330 \\ 
          2012 Qtr2 & 0.18 & 0.18 & 0.00 & 0.93 & 108.14 & 2085 & 1668 \\ 
          2012 Qtr3 & 0.18 & 0.19 & 0.01 & 0.58 & 130.95 & 2494 & 1974 \\ 
          2012 Qtr4 & 0.14 & 0.14 & 0.00 & 0.93 & 137.76 & 2632 & 2067 \\ 
          Female & 0.68 & 0.73 & 0.05 & 0.18 & 89.75 & 1698 & 1422 \\ 
          Clinic A & 0.17 & 0.13 & -0.04 & 0.07 & 100.46 & 1916 & 1579 \\ 
          Clinic B & 0.12 & 0.15 & 0.03 & 0.21 & 108.18 & 2085 & 1668 \\ 
          Clinic C & 0.14 & 0.15 & 0.01 & 0.69 & 114.85 & 2182 & 1755 \\
           \bottomrule
    \end{tabular}
    \begin{tablenotes}[para]
        Note: Each row reports the average effect (at the cutoff) of being assigned to the treatment versus the control condition on a given predetermined covariate. Analysis based on local linear estimation with MSE-optimal bandwidth. The first and second columns report, respectively, the intercepts of the local linear fits to the left and right of the cutoff. The third column, $\widehat{\tau}_{\mathtt{Y}}$, reports the difference between the first two columns, the intention-to-treat RD effect. P-value based on robust bias correction inference methods. The fifth column reports the MSE-optimal bandwidth. Column $N^-_h$ reports the number of observations with score in $[\C-h,\C)$ and column $N^+_h$ reports the number of observations with score in $[\C,\C+h]$.
    \end{tablenotes}
    \end{threeparttable}
}
\end{center}
\end{table}

\subsubsection{Bandwidth Sensitivity, Donut Hole, and Placebo Cutoffs}

This battery of diagnostic tests have all the same underling principle: they investigate the sensitivity of the results to small changes of different features of the implementations and data. The tests consider whether varying the bandwidth or local randomization neighborhood changes the empirical results; whether the observations closest to the cutoff overwhelmingly affect the extrapolation; and whether a fake treatment assignment rule (cutoff) leads to non-zero treatment effects.

The first method focuses on \textit{bandwidth or local randomization neighborhood sensitivity}, which is a common strategy to probe the robustness of the empirical conclusions results to variations of the local neighborhood used for analysis. For example, as discussed previously for continuity-based methods, a larger bandwidth will on average lead to a more precise but also more biased RD treatment effect, if misspecification of the unknown conditional expectations approximations near the cutoff is a concern. To illustrate using the ART application, Table \ref{tab:rdd.hiv1--sens} reports results based on two wider bandwidths (relative to the MSE-optimal choice in Table \ref{tab:rdd.hiv1}); the results show that the conclusions are robust: treatment effects and their associated statistical significance remain qualitatively unchanged. A similar procedure can be used to probe the local randomization window, which can also be shrunk or enlarged to investigate the sensitivity of the main empirical findings; see Table \ref{tab:lr--W}.

\begin{table}[ht]
\begin{center}
\resizebox{\textwidth}{!}{
    \begin{threeparttable}
    \caption{Continuity-Based Sensitivity Diagnostics --- ART Application}\label{tab:rdd.hiv1--sens}
    \begin{tabular}{lccccc}
        \toprule
        \multicolumn{6}{c}{Bandwidth Sensitivity Check$^a$} \\
        \midrule
         & RD Effect & 95\% Robust CI & Bandwidth ($h$) &  $N^-_h$ & $N^+_h$\\ 
          \midrule
        ITT Effect of ART Assignment on ART Initiation & -0.22 & [-0.28,-0.13] & 124.36 & 1634 & 1303 \\ 
          ITT Effect of ART Assignment on Program Retention & -0.15 & [-0.22,-0.05] & 124.36 & 1634 & 1303 \\ 
          Fuzzy Effect of ART Initiation on Program Retention & 0.68 & [0.37,1.00] & 124.36 & 1634 & 1303 \\ 
        \midrule
        ITT Effect of ART Assignment on ART Initiation & -0.23 & [-0.29,-0.14] & 149.36 & 1965 & 1525 \\ 
          ITT Effect of ART Assignment on Program Retention & -0.16 & [-0.23,-0.07] & 149.36 & 1965 & 1525 \\ 
          Fuzzy Effect of ART Initiation on Program Retention & 0.69 & [0.42,0.98] & 149.36 & 1965 & 1525 \\ 
        \midrule  
        \multicolumn{6}{c}{Donut Hole Diagnostic$^b$} \\ 
        \midrule
        ITT Effect of ART Assignment on ART Initiation & -0.23 & [-0.3,-0.14] & 122.24 & 1594 & 1262 \\ 
          ITT Effect of ART Assignment on Program Retention &  -0.14 & [-0.21,-0.04] & 122.24 & 1594 & 1262 \\ 
          Fuzzy Effect of ART Initiation on Program Retention &  0.59 & [0.27,0.89] & 122.24 & 1594 & 1262 \\ 
        \midrule
         \multicolumn{6}{c}{Placebo Cutoffs Diagnostic $c=300$} \\  
        \midrule  
         ITT Effect of ART Assignment on ART Initiation & -0.02 & [-0.19,0.13] & 29.95 & 551 & 547 \\
        ITT Effect of ART Assignment on Program Retention & 0.03 & [-0.14,0.21] & 35.91 & 467 & 442 \\ 
        \midrule
         \multicolumn{6}{c}{Placebo Cutoffs Diagnostic $c=400$} \\
        \midrule
         ITT Effect of ART Assignment on ART Initiation & -0.01 & [-0.10,0.05] & 39.11 & 676 & 571 \\ 
        ITT Effect of ART Assignment on Program Retention & 0.00 & [-0.16,0.18] & 44.13 & 528 & 413 \\ 
        \bottomrule
    \end{tabular}
    \begin{tablenotes}[para]
        Note: The two/three rows in each panel show, respectively, $\widehat{\tau}_{\mathtt{D}}$, $\widehat{\tau}_{\mathtt{Y}}$, and $\widehat{\tau}_{\mathtt{FRD}}$. Analysis based on local linear estimation with MSE-optimal main bandwidth reported in third column. Column labeled ``95\% Robust CI'' reports the robust 95\% confidence intervals based on robust bias-corrected inference. Column $N^-_h$ reports the number of observations with score in $[\C-h,\C)$ and column $N^+_h$ reports the number of observations with score in $[\C,\C+h]$. $^a$Bandwidth used in the sensitivity check are $\pm 10$ relative to the benchmark MSE-optimal bandwidth reported in Table \ref{tab:rdd.hiv1}. $^b$Analysis based on local linear estimation with MSE-optimal main bandwidth reported in third column, but excluding observations with CD4 count equal to 349, 350, and 351.
    \end{tablenotes}
    \end{threeparttable}
}
\end{center}
\end{table}

A related falsification method is the so-called \textit{donut hole sensitivity} method, which is based on the idea that the few observations closest to the cutoff should not drastically determine the empirical results. This is the mirror image of the bandwidth sensitivity: in both cases some observations are included or excluded depending on their score values relative to the cutoff. The donut hole falsification test removes a few observations closest to the cutoff in an attempt to understand the sensitivity of the results to those observations, since polynomial approximations can suffer from biases near the cutoff because of Runge's phenomenon. In practice, this method is easily implemented by using either the continuity-based framework or the local randomization framework, using different subsamples where observations in a symmetric interval around the cutoff are removed, starting with those closest to the cutoff and then progressing with larger intervals around cutoff. No special software is needed beyond the packages used for RD treatment effect estimation and inference (\texttt{rdrobust}, \texttt{rdlocrand}). Importantly, unlike the case of predetermined covariates and placebo outcomes, the same bandwidth or local randomization window used for treatment effect estimation should be used, instead of re-estimating a new bandwidth or window for each new subsample generated by the donut hole. We illustrate the donut hole diagnostic test with the ART application, dropping patients with CD4 count values of $349$, $350$, and $351$ and re-estimating the RD effects using continuity-based methods only to conserve space. We report the results in Table \ref{tab:rdd.hiv1--sens}, which show only minor differences between the donut hole estimates and the main results. This implies that our results are not sensitive to the small set of patients with CD4 counts right around the cutoff.

\begin{table}[ht]
\begin{center}
\resizebox{\textwidth}{!}{
    \begin{threeparttable}
    \caption{Local Randomization Neighborhood Sensitivity Diagnostic --- ART Application}\label{tab:lr--W}
    \begin{tabular}{lcccc}
        \toprule
         & Risk Difference & 95\% Confidence Interval & $N_\W^-$ &  $N_\W^+$ \\ 
         \midrule
          \multicolumn{5}{c}{$\W = [340,360]$}\\
        \midrule
        ITT Effect of ART Assignment on ART Initiation & -0.14 & [-0.25 , -0.04] & 144 & 127 \\ 
          ITT Effect of ART Assignment on Program Retention & -0.09 & [-0.20 , 0.030] & 144 & 127 \\ 
          Fuzzy Effect of ART Initiation on Program Retention & 0.60 & [-0.06 , 1.27] & 144 & 127 \\ 
        \midrule
          \multicolumn{5}{c}{$\W = [335,365]$}\\
        \midrule  
        ITT Effect of ART Assignment on ART Initiation & -0.21 & [-0.30 , -0.13] & 212 & 189 \\ 
          ITT Effect of ART Assignment on Program Retention & -0.11 & [-0.20 , -0.01] & 212 & 189 \\ 
          Fuzzy Effect of ART Initiation on Program Retention & 0.50 & [0.13 , 0.86] & 212 & 189 \\ 
        \midrule
          \multicolumn{5}{c}{$\W = [330,370]$}\\
        \midrule  
        ITT Effect of ART Assignment on ART Initiation & -0.25 & [-0.32 , -0.18] & 277 & 245 \\ 
          ITT Effect of ART Assignment on Program Retention & -0.13 & [-0.21 , -0.06] & 277 & 245 \\ 
          Fuzzy Effect of ART Initiation on Program Retention & 0.52 & [0.25 , 0.79] & 277 & 245 \\  
           \bottomrule
    \end{tabular}
    \begin{tablenotes}[para]
        Note: The three rows in each panel show, respectively, $\widehat{\theta}_{\mathtt{D}}$, $\widehat{\theta}_{\mathtt{Y}}$, and $\widehat{\theta}_{\mathtt{FRD}}$, corresponding to the local randomization estimates based on local randomization window $\W$. Benchmark local randomization window is reported in Table \ref{tab:rdd.hiv1}. Column $N^-_\W$ reports the number of observations with score in $\W$ and below the cutoff ($T_i=0$) and column $N^+_\W$ reports the number of observations with score in $\W$ and above the cutoff ($T_i=1$).
    \end{tablenotes}
    \end{threeparttable}
}
\end{center}
\end{table}

A third sensitivity approach investigates placebo cutoffs using either only control or only treated observations. The idea is to provide evidence in favor of continuity of the regression functions or, more generally, validity of the treatment assignment rule. In a nutshell, this approach analyzes either control or treatment units separately, and sets a sequence of artificial or placebo RD cutoffs to check that there is no RD treatment effect at those alternative cutoffs, since the expectation is that a treatment effect should occur only at the true cutoff and not at artificial cutoffs where treatment status is constant by construction. Empirical evidence of treatment effects at artificial cutoffs may undermine the design if the researcher cannot explain why these effects occur: non-zero effects at artificial cutoffs suggest the possibility that other factors are affecting the units in the background. Table \ref{tab:rdd.hiv1--sens} illustrates the idea with placebo cutoffs $300$ and $400$, using continuity-based methods only to conserve space. For both intention-to-treat effects on program retention and ART initiation, we find that the robust $95\%$ confidence intervals include zero, a reassuring result.

All the empirical results in this section are based on varying arguments in the \texttt{rdrobust} package for continuity-based methods, and in \texttt{rdlocrand} package for local randomization methods, and hence are readily available using general purpose software. See accompanying replication files.

\subsubsection{Fuzzy RD Validation}

To close our discussion of RD falsification methods, we review some validation methods that are specific to fuzzy RD designs. Since the fuzzy RD design shares several features with the IV design, these tests are generally based on diagnostics methods for IV designs. See \cite{Glymour-TchetgenTchetgen-Robins_2012_AJE}, \cite{Baiocchi-Cheng-Small_2014_StatsMed}, \cite{Pizer_2016_HSR}, \cite{Keele-Zhao-Kelz-Small_2018_MC}, and references therein, for reviews and examples of empirical evaluation of IV assumptions in biomedical research and causal inference.

The canonical fuzzy RD estimator is a local version of the standard two-stage least squares estimator in IV settings, and hence it requires a first stage well-separated from zero. Failure of this condition leads to a problem known as ``weak instruments'' in the IV literature, and is a serious concern when analyzing fuzzy RD designs \citep{Feir-Lemieux-Marmer_2016_JBES}. In IV designs, a weak IV test is used to ensure that the effect of the instrument on the treatment exposure is sufficiently strong. The validation analysis of the fuzzy RD design should include this test, with the key difference that the standard weak IV test should be applied within the local neighborhood around the cutoff. Performing the test in a local neighborhood is important: a weak IV test that uses all observations is likely to overstate the strength of the instrument, since it would include data that is excluded from the main analysis by bandwidth or window selection. 

Similarly to the sharp RD design, predetermined covariates should be used to validate the assumptions of the fuzzy RD design. These tests are implemented in the same way as in the sharp RD design, exploring whether the covariates are balanced in a neighborhood of the cutoff. These balance tests should use only the predetermined covariates and the score; the treatment received should not be used for covariate falsification purposes. Another IV diagnostic reports some measure of bias associated with the instrument rather than balance, since bias from a baseline covariate can be amplified by how strongly the IV is predictive of the treatment. This second diagnostic approach can be easily accommodated to the RD setting by applying the fuzzy RD ratio estimator to the baseline covariates (instead of the standard sharp RD estimator) when testing for differences in baseline covariates. See \cite{Davies-etal_2017_IJE} and \cite{branson2020evaluating} for related formal and graphical methods, which we omit to conserve space.

The key assumption known as the exclusion restriction---that being assigned to the treatment has no effect on the outcome except via actual treatment exposure---is untestable. This is a fundamental identifying assumption. For fuzzy RD designs, it is important to evaluate the exclusion restriction using qualitative information. See \citet{Arai-Hsu-Kitagawa-Mourifie-Wan_2022_QE} and references therein for more discussion.

We illustrate these falsification methods with the ART application. We focus on two tests that are specific to RD designs with non-compliance: a weak instrument test, and covariate ``balance'' tests based on the ratio fuzzy RD estimator. We implement a weak IV test in the following way. First, we estimate the MSE-optimal bandwidth for both sides of the cutoff, using ART initiation (the treatment) as the outcome. We then implement a weak IV test using only the observations within this bandwidth. This test consists of regressing the treatment on an indicator variable for whether the CD4 score is less than $350$, and assessing the F-value from this regression. The F-value from the weak IV test is approximately $698$, well above the standard critical value thresholds used in the IV literature. Standard methods can be used to generate these F-values while using bandwidth selection methods in rdrobust; see the replication materials for details. We also test for differences in baseline covariates at the cutoff using the fuzzy RD estimator in Table~\ref{tab:bal2}. The balance results in Table~\ref{tab:cd1} inflate the estimates in Table~\ref{tab:cd1} by the strength of the instrument, which may make it more likely to find imbalances. However, the results are similar to those results in Table~\ref{tab:cd1} based on the intention-to-treat RD estimator: we still find that there are no significant differences in the baseline covariates at the cutoff. 

\begin{table}[ht]
\begin{center}
\resizebox{.9\textwidth}{!}{
    \begin{threeparttable}
    \caption{Continuity-Based Fuzzy RD Estimates for Predetermined Covariates with Robust Bias Corrected Inference --- ART Application}\label{tab:bal2}
    \begin{tabular}{lccccccc}
    \toprule
     & $\widehat{\tau}_{\mathtt{FRD}}$ & Robust p-value & MSE-Optimal Bandwidth & $N^-_h$ & $N^+_h$ \\ 
      \toprule
      Age 0-18 & -0.05 & 0.41 & 122.00 & 2281 & 1835 \\ 
      Age 18 -25 & -0.10 & 0.58 & 91.15 & 1718 & 1440 \\ 
      Age 25-30 & 0.14 & 0.37 & 111.16 & 2093 & 1674 \\ 
      Age 30-35 & 0.01 & 0.88 & 94.99 & 1767 & 1473 \\ 
      Age 35-40 & -0.02 & 0.89 & 110.41 & 2073 & 1660 \\ 
      Age 40-45 & -0.01 & 0.88 & 85.55 & 1592 & 1357 \\ 
      Age 45-55 & 0.00 & 0.90 & 105.77 & 1981 & 1606 \\ 
      Age 55+ & 0.01 & 0.83 & 114.10 & 2138 & 1725 \\ 
      2011 Qtr3 & 0.09 & 0.33 & 112.25 & 2152 & 1717 \\ 
      2011 Qtr4 & 0.09 & 0.54 & 79.37 & 1505 & 1299 \\ 
      2012 Qtr1 & -0.09 & 0.45 & 106.25 & 2034 & 1646 \\ 
      2012 Qtr2 & -0.00 & 0.99 & 85.82 & 1624 & 1383 \\ 
      2012 Qtr3 & -0.05 & 0.64 & 114.33 & 2182 & 1755 \\ 
      2012 Qtr4 & 0.03 & 0.70 & 99.25 & 1891 & 1569 \\ 
      Female & -0.27 & 0.16 & 76.70 & 1443 & 1243 \\ 
      Clinic A & 0.20 & 0.13 & 74.44 & 1416 & 1229 \\ 
      Clinic B & -0.11 & 0.38 & 85.94 & 1624 & 1383 \\ 
      Clinic C & -0.03 & 0.69 & 110.76 & 2116 & 1690 \\ 
       \bottomrule
    \end{tabular}
    \begin{tablenotes}[para]
    Note: The first column is the fuzzy RD effect ($\widehat{\tau}_{\mathtt{FRD}}$) for each predetermined covariate. Analysis based on local linear estimation with MSE-optimal bandwidth. P-value based on robust bias correction inference methods. The third column reports the MSE-optimal bandwidth. Column $N^-_h$ reports the number of observations with score in $[\C-h,\C)$ and column $N^+_h$ reports the number of observations with score in $[\C,\C+h]$.
    \end{tablenotes}
    \end{threeparttable}
}
\end{center}
\end{table}

\section{Analysis with Discrete Score}
\label{sec:RD Analysis with Discrete Score}

When the score only exhibits a ``few'' unique values ($30$ or fewer in typical applications), it is more appropriate to think of the RD design as truly having a discrete score. In that case, the methods presented in the previous sections need to be modified in order to be applicable. Continuity of the score is a key identifying assumption in the continuity-based RD framework, since the continuity assumption is used to extrapolate the information of units near the cutoff under the thought experiment that, in large samples, eventually many units will be arbitrary close to the cutoff. Although continuity-based methods can be used under parametric assumptions when the score exhibits repeated values or mass points in its distribution, the local randomization RD framework is more naturally applicable in this setting because, by construction, this framework assumes valid extrapolation within the local randomization neighborhood---no futher parametric assumptions are needed. Furthermore, the local randomization framework allows for employing only the closest observations to the cutoff, which tends to minimize extrapolation biases. In this section, we discuss how to employ both RD frameworks when the score is discrete with a few mass points, focusing only on the changes in interpretation and implementation that are required.

We consider a second biomedical empirical illustration on patient cost-sharing and healthcare utilization. In many countries, health care costs are subsidized through government programs, and research in health policy and management seeks to understand whether lower levels of cost-sharing encourages patients to use healthcare services at higher rates. Government health care cost-sharing often varies by age, thereby creating a discontinuity in health care subsidization that can be analyzed using RD methods. In this type of RD design, researchers compare health care utilization for those just above and below the age at which cost-sharing levels change. For example, in the U.S., eligibility for the healthcare program Medicare starts at age $65$, and thus a common RD empirical strategy compares health care usage or other outcomes of interest for adults just above and below this age threshold. We reanalyze the recent work of \citet{han2020patient}, who studied patient cost-sharing and healthcare utilization in Taiwan, where all inpatient and outpatient services for children under the age of $3$ are exempt from co-payments. The authors used this discontinuity in age to compare levels of health care utilization just before and after the third birthday. Henceforth, we refer to this study as the \textit{cost-sharing} application.

The raw data includes $414,282$ children born between $2003$ and $2004$, and records the number of days until the child's third birthday---normalized to be zero on the day of the third birthday. Data on healthcare utilization was collected for up to $180$ days before and after each child's third birthday. The treatment is an indicator equal to $1$ if the child's age at the time of their visit is $3$ or greater. This captures the higher level of patient’s cost-sharing due to the expiration of the subsidy after the third birthday. The outcome of interest is the number of medical visits per $10,000$ person days, and thus the outcome variable $Y_i$ records the average for all children within a day relative to the cutoff. The data was aggregated at the day-level, but the score records only the associated week for each day, with a total of $50$ weeks. While the unit of analysis is at the day-level, the score variable records the week relative to the cutoff, so there are seven observations for each value of the score $X_i\in\{-25,-24,\dots,0,\dots,23,24\}$. Unlike the ART application, this RD design is sharp because once the child is $3$ years of age or older there are no exceptions to the change in cost-sharing. 

\begin{figure}[ht]
    \centering
	\begin{subfigure}{0.48\textwidth}
		\centering
		\includegraphics[scale=.45]{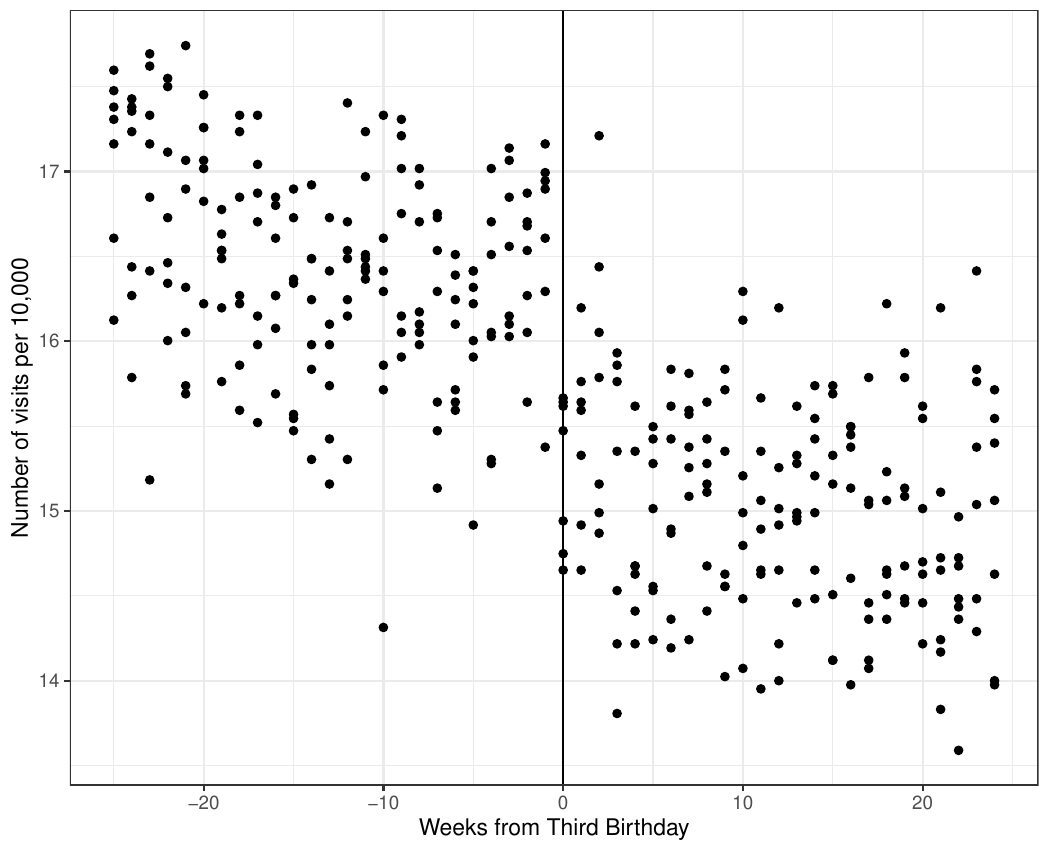}
		\caption{Scatter plot: Outcome ($Y_i$) vs. Score ($X_i$)}\label{fig: Cost-Sharing - Scatter plot}
	\end{subfigure}
	\begin{subfigure}{0.48\textwidth}
		\centering
		\includegraphics[scale=0.45]{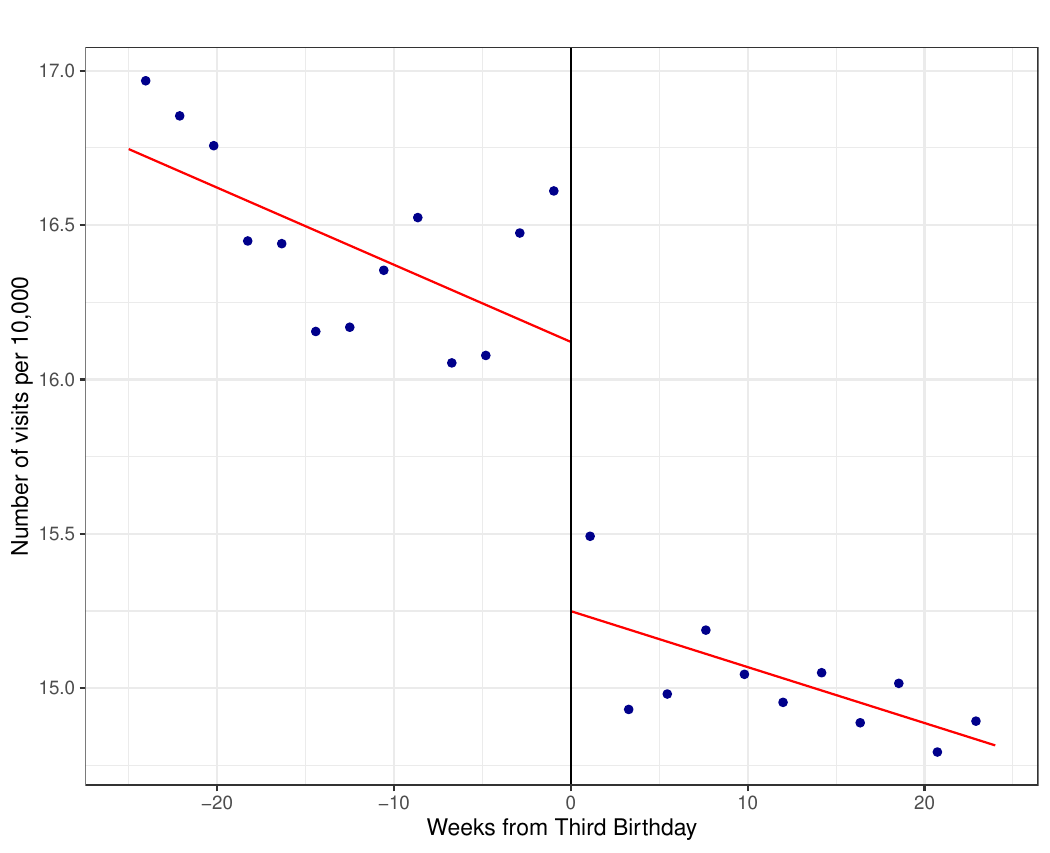}
		\caption{RD Plot: Outcome ($Y_i$) vs. Score ($X_i$)}\label{fig: Cost-Sharing - RD plot}		
	\end{subfigure}
	\caption{Basic Plots --- Cost-Sharing Application}\label{fig: Cost-Sharing - Basic plots}
   \flushleft{{\footnotesize Note: The score $X_i$ is the number of weeks until the child turns five years of age, where the date of birth is normalized to zero, so that positive numbers represents weeks past the third birthday, and negative numbers represent weeks before the third birthday. The cutoff is zero. Children below the cutoff are eligible for a health care subsidy; the subsidy is not available for children above the cutoff. The outcome $Y_i$ is the number of medical visits per 10,000 person days. Both panels display the outcome against the score. In panel (b), dots are local means of $Y_i$ calculated in different non-overlapping bins of $X_i$, and the solid line is a 4th-order polynomial of $Y_i$ on $X_i$, fitted separately for patients above and below the cutoff.}}
\end{figure}

Graphical presentation continues to be important when the score is discrete. However, in this case, histograms, scatterplots, and RD plots can be constructed directly by employing the unique values of the score; there is no need for binning the score data first as in Section \ref{sec:Analysis with Continuous Score}. In the case of the cost-sharing application, the histogram would not be informative since each mass point contains exactly seven observations. Thus, Figure \ref{fig: Cost-Sharing - Basic plots} presents a scatter plot and a RD plot instead.

Figure \ref{fig: Cost-Sharing - Scatter plot} shows that indeed seven observations share each unique value of the score. Furthermore, the scatter plot suggests that  when children are three or older, the rate of hospital visits drops. Figure \ref{fig: Cost-Sharing - RD plot} presents an RD plot, where each bin is equivalent to one value of the score, so each dot reports the sample average of the seven observations for that week. More generally, if the score variable takes on the values $\{\mathsf{x}_{K_{-}}, \dots, \mathsf{x}_{-2},\mathsf{x}_{-1},\C,\mathsf{x}_1,\mathsf{x}_2,\dots,\mathsf{x}_{K_{+}}\}$, with $K_-$ denoting the number of unique values below the cutoff and $K_+$ denoting the number of unique values above the cutoff, then the RD plot reports $K=K_- + K_+ + 1$ sample averages of the outcome for each score value, $\{\bar{Y}_{K_{-}}, \dots,\bar{Y}_{-2},\bar{Y}_{-1},\bar{Y}_\C,\bar{Y}_1,\bar{Y}_2,\dots,\bar{Y}_{K_{+}}\}$, where $\bar{Y}_{j} = \frac{1}{\# \{i: x_i = \mathsf{x}_j\}}\sum_{i: x_i = \mathsf{x}_j} Y_i$ for $j \in \{ K_{-}, \ldots, -2,-1,\C,1,2,\ldots, K_{+}\}$ and $\#\{A\}$ denotes the number of elements in the set $A$. The global polynomial fits are added for visual presentation only, as they represent a global parametric interpolation across the $K$ unique values or mass points available for the score. 

\subsection{Continuity-Based Methods}

In RD settings where there are repeated score values among the units, continuity-based methods can be applied if the score takes on a relatively large number of distinct values and the researcher is willing to make additional assumptions. A reasonable approach is to view the number of unique values of the score $K$ as the effective sample size, and thus treat the units with identical scores as independent measurements of each particular score level. From this perspective, the total sample size continues to be $n$ but the effective sample size is smaller because there is only $K \leq n$ unique values among $X_1,X_2,\dots,X_n$. 

Whenever the number of the unique score values $K$ is large enough and the closest mass points to the cutoff are close enough, the continuity-based framework can be taken as a reasonable approximation for identification purposes and hence $\tau_\mathtt{SRD}$, $\tau_\mathtt{Y}$, $\tau_\mathtt{D}$ and $\tau_\mathtt{FRD}$ are reasonable treatment effects of interest. The key requirement is that whatever extrapolation takes place from the closest observed score value to the cutoff has a sufficiently small error. Clearly, some extrapolation would be needed, which is ultimately achieved via the estimation and inference methods employed. 

Local polynomial methodology can be adapted and used for both optimal point estimation and robust bias corrected inference in settings with a large number of unique score values. The only important change is related to the sample size used, in addition to the key identifying assumptions invoked. From an approximation error perspective, only variation in the score variable can reveal the shape of the conditional expectation functions, and hence the correct sample size to be considered is $K$, not $n$. On the other hand, from an uncertainty perspective, either $K$ or $n$ could be the correct sample size, depending on the assumptions imposed about how the data was generated. Either way, the presence of repeated score values in the sample affects bandwidth selection and inference methods, but the necessary modifications are straightforward and readily applicable in general purpose software (\texttt{rdrobust} and \texttt{rddensity} packages). This logic justifies the empirical analysis in Section \ref{sec:Analysis with Continuous Score}, where we used continuity-based methods despite having some repeated score values, because $K$ was large. See \citet[Section 3]{Cattaneo-Idrobo-Titiunik_2023_Book} and references therein.

The situation is different when $K$ is small, that is, when there are only a few unique values in the score (usually $30$ or less). In this case, it may be unreasonable to assume valid nonparametric extrapolation to the cutoff because it would be hard (or impossible) to learn the shape of the conditional expectation functions arbitrarily close to the cutoff. In this scenario, a solution to validate continuity-based methods is to rely on parametric extrapolation, where by virtue of the coarseness of the score, the postulated local polynomial model must be assumed to be correctly specified. This is a strong assumption, but necessary to restore point identification of RD treatment effect parameters at the cutoff. 

To illustrate the point with an extreme example, suppose the score $X_i$ takes only on five distinct values $\mathsf{x}_{-2}<\mathsf{x}_{-1}<\C<\mathsf{x}_1<\mathsf{x}_2$, where $\C$ continues to denote the RD cutoff. It follows that only $\E[Y_i(0) | X_i=\mathsf{x}_{-2}]$, $\E[Y_i(0) | X_i=\mathsf{x}_{-1}]$, $\E[Y_i(1) | X_i=\C]$, $\E[Y_i(1) | X_i=\mathsf{x}_1]$ and $\E[Y_i(1) | X_i=\mathsf{x}_2]$ are identifiable from the data. In particular, $\tau_{\mathtt{SRD}} = \E[Y_i(1) | X_i=\C] - \E[Y_i(0) | X_i=\C]$ will never be nonparametrically identifiable because $\E[Y_i(0) | X_i=\C]$ is not identifiable without parametric assumptions about the functional form of $\E[Y_i(0) | X_i=x]$ for $x\in(\mathsf{x}_{-1},\C]$. Moreover, $\E[Y_i(1) | X_i=\C]$ will be nonparametrically identifiable in a super-population sense only in settings where $\mathbb{P}[X_i=\C]>0$, that is, when the number of repeated values at $X_i=\C$ is sufficiently large. This example shows a more general phenomenon: if the score is discrete, the canonical continuity-based RD parameters $\tau_\mathtt{SRD}$, $\tau_\mathtt{Y}$, $\tau_\mathtt{D}$, or $\tau_\mathtt{FRD}$ are not point identifiable without strong, parametric assumptions about the functional form of $\E[Y_i(0) | X_i=x]$ and $\E[Y_i(1) | X_i=x]$. This leads to two possible conceptual approaches: (i) assume such parametric assumptions hold, or (ii) change the parameter of interest. As already discussed, continuity-based RD methods can be deployed to RD designs with discrete score variables whenever the local parametrizations are assumed to generate small misspecification bias, that is, when the local polynomial model is assumed to be approximately correctly specified.

We illustrate these ideas with the cost-sharing application. The RD effect is $-1.29$ with robust confidence interval of $[-2.08,-0.75]$ and robust p-value of zero (main MSE-optimal bandwidth equal to $6.08$ weeks). See Table~\ref{tab:taiwanCresults} for the full results.

\begin{table}[ht]
\begin{center}
\resizebox{\textwidth}{!}{
    \begin{threeparttable}
    \caption{Continuity-Based Sharp RD Methods --- Cost-Sharing Application}\label{tab:taiwanCresults}
    \begin{tabular}{lccccc}
        \toprule
         & RD effect & 95\% Robust CI & Bandwidth ($h$) &  $N^-_h$ & $N^+_h$\\ 
        \midrule
         Number of Doctor Visits per 10,000 & -1.29 & [-2.08,-0.75] & 6.08 & 42 & 49 \\  
        \bottomrule
    \end{tabular}
    \begin{tablenotes}[para]
        Note: Analysis based on local linear estimation with MSE-optimal main bandwidth reported in third column. Column labeled ``95\% Robust CI'' reports the robust 95\% confidence intervals based on robust bias-corrected inference. Column $N^-_h$ reports the number of observations with score in $[\C-h,\C)$ and column $N^+_h$ reports the number of observations with score in $[\C,\C+h]$.
    \end{tablenotes}
    \end{threeparttable}
}
\end{center}
\end{table}

\subsection{Local Randomization Methods}

Provided the parameter of interest is changed or reinterpreted appropriately, RD identification, estimation and inference under a local randomization framework remains valid when the score exhibits mass points. To formalize the core ideas, we continue to assume that the support of the score variable is $\{\mathsf{x}_{K_{-}}, \dots, \mathsf{x}_{-2},\mathsf{x}_{-1},\mathsf{x}_\C,\mathsf{x}_1,\mathsf{x}_2,\dots,\mathsf{x}_{K_{+}}\}$, with $K= K_{-} + K_{+} + 1$ the total number of unique values. The local randomization assumption reduces to specifying a window containing some of these unique values where the two local randomization conditions discussed in Section \ref{sec:RDgeneral} are assumed to hold.

We can define the following alternative RD parameters for settings where the score has few mass points: $\tilde{\theta}_\mathtt{SRD} = \E[Y_i(1) | X_i=\C] - \E[Y_i(0) | X_i=\mathsf{x}_{-1}]$, $\tilde{\theta}_\mathtt{Y} = \E[Y_i(1,D_i(1)) | X_i=\C] - \E[Y_i(0,D_i(0))| X_i=\mathsf{x}_{-1}]$, $\tilde{\theta}_\mathtt{D} = \E[D_i(1) | X_i=\C] - \E[D_i(0) | X_i=\mathsf{x}_{-1}]$, and $\tilde{\theta}_\mathtt{FRD} = \tilde{\theta}_\mathtt{Y}/\tilde{\theta}_\mathtt{D}$. The notation makes clear that the parameters of interest have changed: they now correspond to comparisons of potential outcomes at different values of the score variable ($X_i=\C$ vs. $X_i=\mathsf{x}_{-1}$). This approach allows for the deployment of local randomization RD methods. First, because the Fisherian approach is finite-sample valid, this method can be used even with small sample size at the two score evaluation points $X_i=\C$ and $X_i=\mathsf{x}_{-1}$. The super-population approach, in contrast, relies on large sample approximations and consequently requires a large enough number of repeated values at $X_i=\C$ and $X_i=\mathsf{x}_{-1}$. In practice, this idea can be used for the two closest values to the cutoff or, alternatively, for a collection of unique values closest to the cutoff. As before, the number of unique points on the score closest to the cutoff used is determined by the choice of window $\W$.

The choice of $\W$ in this case is simplified considerably. The implementation of the window selector based on covariates should start with the smallest possible window,  $[\mathsf{x}_{-1}, \C]$, and continue increasing this window one mass point at a time on either side. If there are enough observations in  the window $[\mathsf{x}_{-1}, \C]$, researchers should report results for this window. Even if a larger window is chosen by the covariate-based window selector, it will be important to show the results when only the observations closest to the cutoff are included in the analysis. Whenever $\W$ contains enough unique values of the score, it is also possible to use parametric extrapolation ideas. In this case, a parametric relationship is postulated between the outcome variables and the score, and regression-based methods are used for adjustment.

We illustrate these ideas with the cost-sharing application. We use the local randomization approach and implement a window selector to find the largest window around the cutoff where all covariates are balanced in that window and in all the windows contained in it. We use four pre-determined covariates in our window selector: share of male children, household income per capita, share of children born in Taipei, and birth year. The results show that only the first window, which has seven observations on each side, has all covariates balanced; starting in the second window (14 days on either side of the cutoff), the minimum p-value is well under 5\% (in fact, it is zero for all windows after the second). Table \ref{tab:taiwanwin1} shows the results of the balance tests in our selected window, where not only are the p-values above a $0.15$ threshold, but the differences in means are very small for all four covariates. 

\begin{table}[ht]
\begin{center}
\resizebox{.9\textwidth}{!}{
    \begin{threeparttable}
    \caption{Distribution of Predetermined Covariates for $\W=[-1,0]$ --- Cost-Sharing Application}\label{tab:taiwanwin1}
    \begin{tabular}{lcccc}
    \toprule
     & Mean Below & Mean Above & Diff. in Means & p-value \\ 
    \midrule
      Share of male & 0.55 & 0.55 & 0.00 & 0.78 \\ 
      Household Income per Capita & 12494.53 & 12532.86 & 38.33 & 0.21 \\ 
      Share of children born in Taipei & 0.08 & 0.08 & 0.00 & 0.80 \\ 
      Birth year & 2003.49 & 2003.49 & -0.00 & 0.21 \\ 
    \bottomrule
    \end{tabular}
    \begin{tablenotes}[para]
        Note: Only including children who are within $7$ days of their third birthday; there are $14$ total observations, $7$ on each side of the cutoff. The last column shows the Fisherian p-value assuming a fixed margins randomization mechanism that assigns these $14$ observations to be above or below the birthday cutoff (which is normalized at zero).
    \end{tablenotes}
    \end{threeparttable}
}
\end{center}
\end{table}

We now estimate the effect of the treatment in the selected window. Since the number of observations in this window is only $14$, it is important that we use Fisherian methods for inference, since those do not rely on large-sample approximations and provide exact p-values even when sample sizes are very small as in this case. The results are shown in Table \ref{tab:taiwanYresults}, where we see that the mean difference in the number of doctor visits per $10,000$ is $-1.362$: the number of visits per $10,000$ is $16.61$ for children who are two years old and whose third birthday is within $7$ days, compared to $15.248$ for children who turned three in the past seven days. The Fisherian p-value associated with the test of the hypothesis that there is no effect for any unit is well below $1\%$, despite the low sample size. This suggests, as expected, that cost sharing causes families to use medical care at higher rates.

\begin{table}[ht]
\begin{center}
\resizebox{.9\textwidth}{!}{
    \begin{threeparttable}
    \caption{Local Randomization Sharp RD Methods --- Cost-Sharing Application}\label{tab:taiwanYresults}
    \begin{tabular}{lcccc}
    \toprule
      \multicolumn{5}{c}{Local Randomization Estimate} \\ 
    \midrule
     & Mean Below & Mean Above & Diff. in Means & Fisherian p-value \\ 
    \midrule
     Number of Doctor Visits per 10,000& 16.610 & 15.248  &-1.362 &0.006  \\ 
     \midrule
    \multicolumn{5}{c}{Local Randomization Placebo Estimate (Pre-treatment Period)} \\ 
    \midrule
     Number of Doctor Visits per 10,000&  11.918 & 11.987 & 0.069 & 0.720  \\ 
    \bottomrule
    \end{tabular}
    \begin{tablenotes}[para]
        Note: Only including children who are within $7$ days of their third birthday; there are $14$ total observations, $7$ on each side of the cutoff. The last column shows the Fisherian p-value assuming a fixed margins randomization mechanism that assigns these $14$ observations to be above or below the birthday cutoff (which is normalized at zero).
    \end{tablenotes}
    \end{threeparttable}
}
\end{center}
\end{table}

Finally, we conduct a placebo analysis to assess the validity of the design. The data contains information on healthcare utilization for the period from 1997-2002, when cost-sharing was not higher for children under three. Thus, we expect no treatment effects from a similar analysis in this pre-treatment period. Indeed, Table \ref{tab:taiwanYresults} shows that the Fisherian sharp null of no treatment effect cannot be rejected, with a p-value of $0.209$. Moreover, the difference in means of $-0.127$ is less than one tenth of the $-1.362$ difference observed in the post-treatment period. Similar null results are obtained if the window in the placebo analysis is widened to plus or minus four weeks of the date of the third birthday.

\subsection{Evaluating the RD Assumptions}

In Section \ref{sec:Analysis with Continuous Score}, we discussed an array of falsification and validation methods for RD designs with a continuous score variable. Those methods can be directly employed in settings where the score is discrete but the number of unique score values $K$ is large enough. When $K$ is small, some of these methods are easily applicable while others are not. For the \textit{score density near the cutoff} diagnostic, the binomial test continues to be valid regardless of the size of $K$ because this is a finite-sample valid test about the relative proportion of units on either side of the cutoff. On the other hand, the density test must be handled with more care because that method was developed for (approximately) continuously distributed scores. For \textit{predetermined covariates and placebo outcomes} diagnostics, all ideas and methods discussed in this section can be applied directly. \textit{Bandwidth sensitivity} diagnostics can be applied when the score is discrete, while the \textit{donut hole} and \textit{placebo cutoff} diagnostic are more difficult to implement without strong parametric assumptions. Finally, \textit{fuzzy RD validation} diagnostics can be adapted from the standard IV literature straightforwardly.

\subsection{A Flawed RD Application: Genetic Assay Guidelines for Chemotherapy}

The ART and cost-sharing applications showcase RD designs with many and with few discrete score values that pass all the key diagnostic tests, and produce robust and statistically significant treatment effect estimates. To contrast, we now discuss a third empirical application where the key falsification methods do not support the use of RD methods.

While treatment options for breast cancer have greatly expanded over the last two decades, chemotherapy is still often indicated for patients. To guide whether chemotherapy should be administered there are several commercially available gene-expression assays that provide prognostic information in hormone-receptor positive breast cancer patients. One widely used score is the Oncotype DX by Genomic Health, which is a $21$-gene recurrence-score assay that ranges from $0$ to $100$ and is predictive of chemotherapy benefit when it is high---with a high score defined as 31 or higher. When the oncotype score is low ($0$ to $10$), it is prognostic for a very low rate of distant breast cancer recurrence ($2$\%) and adjuvant chemotherapy is not recommended. There is, however, considerable uncertainty as to whether chemotherapy is beneficial for patients who have a mid-range oncotype score. Current clinical guidelines suggest initiation of adjuvant chemotherapy for patients with an oncotype score of $26$ or higher \citep{Paik-etal_2006_JCO,Sparano-Paik_2008_JCO,Albain-etal_2010_Lancet}. Thus, this setup suggests the use of an RD design with a discrete running variable being the oncotype score and the cutoff is $26$.

For this application, we analyze a cohort of patients from the Penn Breast Database from $2009$ to $2017$ with oncotype scores of less than $40$ who underwent surgery and were then eligible for adjuvant chemotherapy. Excluding patients with oncotype scores of $40$ or greater reduces the cohort from $16,488$ to $3,269$, after also excluding $3$ patients who did not undergo oncotype scoring. The database includes several predetermined covariates: age, race, tumor size, tumor grade, an indicator for lymphovascular invasions, an indicator for estrogen receptor, an indicator for progesterone receptor, type of surgery (mastectomy or breast conservation), and an indicator for endocrine therapy. This is an RD design where the unit of observation is the patient, $X_i$ is the oncotype score, $c=26$, the treatment is the receipt of adjuvant chemotherapy, and the outcome of interest is an indicator for recurrence of breast cancer. This RD design is fuzzy since adjuvant chemotherapy was only prescribed to patients. Henceforth, we refer to this empirical application as the \textit{chemotherapy} application.

\begin{figure}[ht]
    \centering
	\begin{subfigure}{0.48\textwidth}
		\centering
		\includegraphics[scale=0.45]{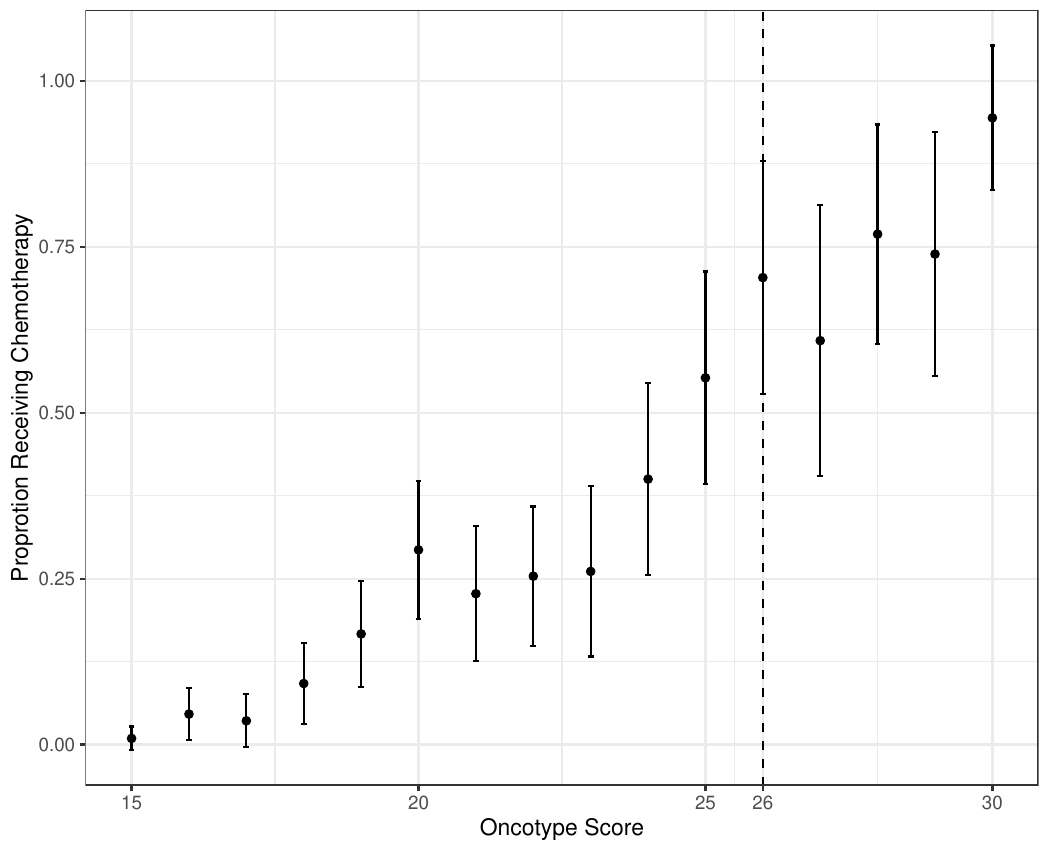}
		\caption{RD Plot: Outcome ($Y_i$) vs. Score ($X_i$) with $95\%$ asymptotic confidence intervals}\label{fig.rf2}		
	\end{subfigure}
	\begin{subfigure}{0.48\textwidth}
		\centering
		\includegraphics[scale=.45]{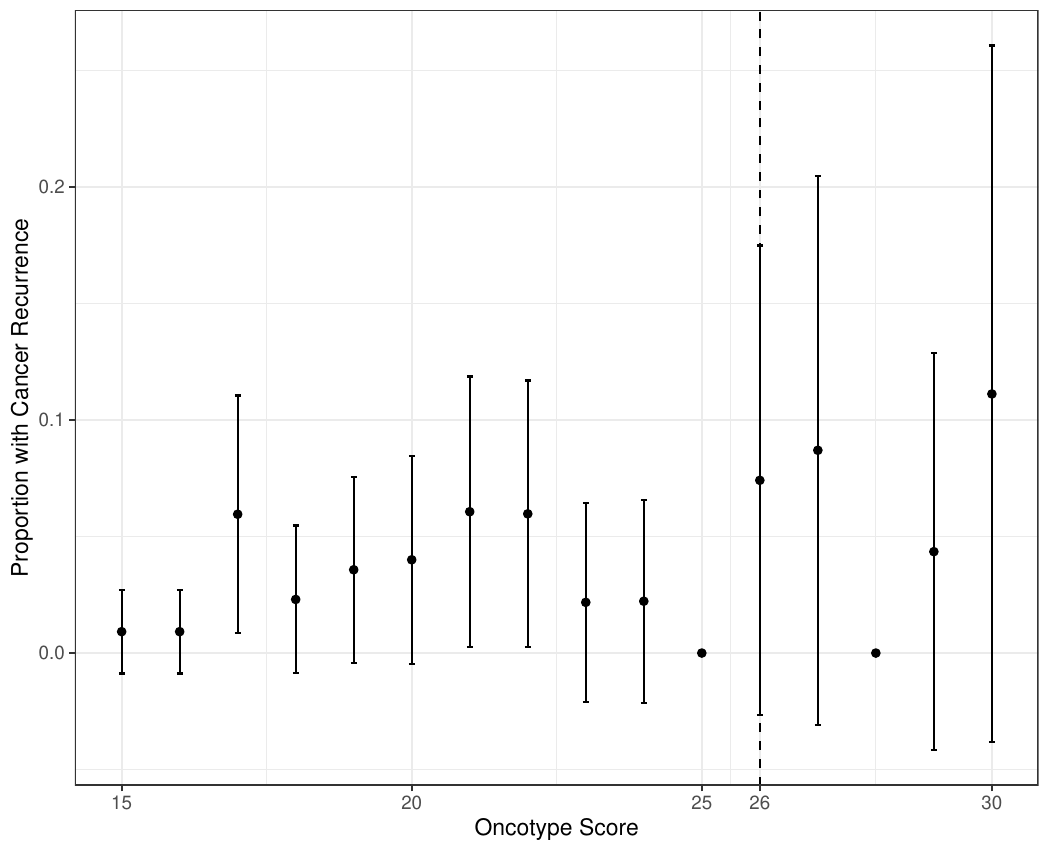}
		\caption{RD Plot: Treatment Take-up ($D_i$) vs. Score ($X_i$)  with $95\%$ asymptotic confidence intervals}\label{fig.rf1}
	\end{subfigure}
	\caption{Basic Plots --- Chemotherapy Application}\label{fig: Chemotherapy Application}
    \flushleft{{\footnotesize Note: The score $X_i$ is the Oncotype DX score for patient $i$. The cutoff is 26: the guideline is to initiate adjuvant chemotherapy for
patients with a score of 26 or higher. The outcome $Y_i$ is an indicator for recurrence of breast cancer. Panel (a) plots the proportion of patients receiving treatment against the score; panel (b) reports the outcome against the score. In both panels, the dots are sample means and the bars represent 95\% confidence intervals.}}
\end{figure}

Figure \ref{fig: Chemotherapy Application} illustrates the fuzzy RD design. Figure \ref{fig.rf2} shows the plot of the breast cancer recurrence indicator on the oncotype score, while Figure \ref{fig.rf1} shows the treatment take-up. Since this score takes only a smaller number of values, this is an example of an RD design with discrete score. We thus simply plot the proportion of patients with breast cancer recurrence for each one of these values. There is an increase in the proportion of patients that receive chemotherapy at the $26$ cutoff, but not all patients with oncotype score of $26$ receive it, and a considerable share of patients with oncotype scores below $26$ are treated with chemotherapy. Moreover, the proportion of treated patients ``jumps'' not only at $26$, but also at $25$ and at $24$, suggesting that some physicians are using a cutoff that is lower than the guideline, or perhaps not using a cutoff at all and simply steadily increasing the probability of chemotherapy treatment as the oncotype score increases. The evidence thus shows that many physicians initiate treatment for patients that are below the clinical guideline that is the basis for the RD design. In general, this pattern will tend to occur in applications where the physician deems the side effects of treatment to be small but the effects of treatment worthwhile. As we demonstrate below, this phenomenon will tend to make the instrument weak in the fuzzy RD design, and preclude the researcher's ability to learn about the treatment effect.

We attempt to validate the RD design by analyzing whether the number of treated and control observations is similar in a small neighborhood of the cutoff, by studying whether these observations are similar in terms of predetermined characteristics or covariates, and by investigating the presence of weak instruments. We implement the density test by applying the binomial test to the oncotype data. For the neighborhood of $W=[25,26]$, there are $38$ observations below the threshold and $27$ above the threshold; assuming the null probability is 1/2, the $p$-value from the binomial test is $0.215$. If instead we use the neighborhood $W=[24,27]$, there are $83$ observations with $50$ above the threshold for a $p$-value of less than $0.005$. The latter result is not consistent with a Bernoulli trial with probability of success $1/2$. As oncotype scores increase, they become less common, and as such there is a clear downward trend in the density of the score. This results in a statistically significant imbalance in the number of units below and above the cutoff as soon as the second largest window is considered.

Next, we explore whether there is a window around the cutoff where all the covariates are balanced. We summarize the results without reporting details to conserve space. Considering the p-value for differences-in-means obtained for each increasing symmetric window from $[25,26]$ to $[21,30]$, we find that for the smallest window the $p$-value for one covariate, tumor size, is less than $0.05$. If tumor size affects cancer recurrence, an imbalance in this covariate can invalidate the outcome comparisons between the treated and the control groups. Ideally, no important confounders should be imbalanced in the chosen local randomization window. Furthermore, the minimum p-value is below $0.10$ in all of these windows, showing that covariate balance is getting worse as the window size increases---a pattern that is expected when the score correlates strongly with units' characteristics. We also found a second covariate, lymphovascular invasion, that is imbalanced in the second smallest window. These pre-intervention covariates are likely to be important determinants of the outcome, and thus the outcome of the RD methods are likely to be confounded and fail to provide a valid estimate of the true effect of chemotherapy on cancer recurrence. In sum, we find statistically significant differences in key pre-intervention features that are likely to confound the RD design. 

Finally, we consider the first-stage treatment effect, $\theta_\mathtt{D}$, to understand whether reaching an oncotype score of $26$ resulted in a significant increase in the probability of being treated with adjuvant chemotherapy. For the smallest window, $\W=[25,26]$ with $N_\W = 65$, we find $\widehat{\theta}_\mathtt{D}=0.15$ with a Fisherian p-value of $0.32$. In addition, the large-sample F-statistic is $1.51$, suggesting a clear problem of weak instruments, as already anticipated in Figure \ref{fig.rf2}. Reaching the cutoff of 26 does not seem to have induced an increase in the probability of receiving adjuvant chemotherapy. Fuzzy RD treatment effects are weakly identifiable and thus unreliable in this application.

In sum, the \textit{chemotherapy} application does not pass basic RD validation/diagnostic tests, and the evidence does not support an RD analysis. The clinical guideline was not followed closely, which resulted in a very weak instrument, and the fact that important confounders are imbalanced even in the smallest windows around the cutoff makes this application an example of a flawed RD design.

\section{Conclusion}\label{sec:conclusion}

The RD design offers biomedical researchers the possibility of rigorously studying the effect of a treatment that is assigned based on a score and a cutoff, such as a recommendation to treat patients with a diagnostic laboratory test above or below a given threshold. Although the patient population above the cutoff cutoff will typically be very different from the patient population below the cutoff, the RD design restores comparability by focusing on patients whose scores are close to the cutoff on either side.

Our discussion, empirical examples, and accompanying computer code provides a state-of-the-art introduction and practical guide for the analysis of canonical RD designs in biomedical contexts. We covered modern estimation, inference, validation, and visualization approaches for the analysis and interpretation of RD designs based on both continuity-based and local randomization frameworks. The main takeaways are as follows: (i) always employ graphical approaches to uncover basic features of the design but never for formal estimation and inference; (ii) always localize near the cutoff when deploying principled statistical methods and never rely on global estimation and inference approaches; (iii) depending on the coarseness of the score, employ continuity-based or local randomization methods being cognizant of the unavoidable extrapolation (to the cutoff) underlying the methods (e.g., ART and Cost-Sharing applications); (iv) always employ falsification and validation diagnostics to offer evidence in favor of the RD design; and (v) be aware that not every treatment assignment mechanism based on a hard-thresholding rule can be analyzed via RD designs methods because sometimes the key underlying identifying assumptions do not hold (e.g., Chemotherapy application).

The canonical RD design can be generalized to the case of multiple scores \citep{PapayWillettMurnane2011-JoE,ReardonRobinson2012-JREE}, the geographic RD design \citep{Keele-Titiunik_2015_PA,Keele-Titiunik-Zubizarreta_2015_JRSSA}, multiple cutoffs \citep{Cattaneo-Keele-Titiunik-VazquezBare_2016_JOP,Cattaneo-Keele-Titiunik-VazquezBare_2021_JASA}, boundary discontinuity designs \citep{Cattaneo-Titiunik-Yu_2023_wp}, kink RD designs \citep{Card-Lee-Pei-Weber_2015_ECMA}, RD designs with rounded scores \citep{Barreca-Lindo-Waddell_2016_EconInq}, and RD designs with measurement error \citep{Bartalotti-Brummet-Dieterle_2021_JBES}, among many other possibilities. We do not discuss all these extensions and generalization due to space constraints. See \citet{Cattaneo-Titiunik_2022_ARE} for more examples and references, and \citet{Cattaneo-Idrobo-Titiunik_2020_Book,Cattaneo-Idrobo-Titiunik_2023_Book} for practical introductions.

All the methods discussed in this tutorial employ open source general-purpose software for \texttt{R}, \texttt{Stata} and \texttt{Python}, available at \url{https://rdpackages.github.io/}. We also provide full replication materials (data and codes) for the three applications, available at \url{https://rdpackages.github.io/replication/}.

\onehalfspacing

\bibliographystyle{jasa}
\bibliography{Cattaneo-Keele-Titiunik_2023_SIM}

\begin{thebibliography}{73}
\newcommand{\enquote}[1]{``#1''}
\expandafter\ifx\csname natexlab\endcsname\relax\def\natexlab#1{#1}\fi

\bibitem[Abadie and Cattaneo(2018)]{Abadie-Cattaneo_2018_ARE}
Abadie, A., and Cattaneo, M.~D. (2018), \enquote{Econometric Methods for
  Program Evaluation,} \emph{Annual Review of Economics}, 10, 465--503.
\bibitem[Albain {\normalfont et~al.}(2010)Albain, Barlow, Shak, Hortobagyi,
  Livingston, Yeh, Ravdin, Bugarini, Baehner, Davidson {\normalfont
  et~al.}]{Albain-etal_2010_Lancet}
Albain, K.~S., Barlow, W.~E., Shak, S., Hortobagyi, G.~N., Livingston, R.~B.,
  Yeh, I.-T., Ravdin, P., Bugarini, R., Baehner, F.~L., Davidson, N.~E.
  {\normalfont et~al.} (2010), \enquote{Prognostic and Predictive Value of the
  21-gene Recurrence Score Assay in Postmenopausal Women with Node-Positive,
  Oestrogen-Receptor-Positive Breast Cancer on Chemotherapy: A Retrospective
  Analysis of A Randomised Trial,} \emph{The Lancet Oncology}, 11, 55--65.
\bibitem[Arai {\normalfont et~al.}(2022)Arai, Hsu, Kitagawa, Mourifi{\'e} and
  Wan]{Arai-Hsu-Kitagawa-Mourifie-Wan_2022_QE}
Arai, Y., Hsu, Y., Kitagawa, T., Mourifi{\'e}, I., and Wan, Y. (2022),
  \enquote{Testing Identifying Assumptions in Fuzzy Regression Discontinuity
  Designs,} \emph{Quantitative Economics}, 13, 1--28.
\bibitem[Arai and Ichimura(2018)]{Arai-Ichimura_2018_QE}
Arai, Y., and Ichimura, H. (2018), \enquote{Simultaneous Selection of Optimal
  Bandwidths for the Sharp Regression Discontinuity Estimator,}
  \emph{Quantitative Economics}, 9, 441--482.
\bibitem[Arai {\normalfont et~al.}(2021)Arai, Otsu and
  Seo]{Arai-Otsu-Seo_2021_wp}
Arai, Y., Otsu, T., and Seo, M.~H. (2021), \enquote{Regression Discontinuity
  Design with Potentially Many Covariates,} \emph{\emph{arXiv:2109.08351}}.
\bibitem[Baiocchi {\normalfont et~al.}(2014)Baiocchi, Cheng and
  Small]{Baiocchi-Cheng-Small_2014_StatsMed}
Baiocchi, M., Cheng, J., and Small, D.~S. (2014), \enquote{Instrumental
  variable methods for causal inference,} \emph{Statistics in Medicine}, 33,
  2297--2340.
\bibitem[Barreca {\normalfont et~al.}(2016)Barreca, Lindo and
  Waddell]{Barreca-Lindo-Waddell_2016_EconInq}
Barreca, A.~I., Lindo, J.~M., and Waddell, G.~R. (2016),
  \enquote{Heaping-Induced Bias in Regression-Discontinuity Designs,}
  \emph{Economic Inquiry}, 54, 268--293.
\bibitem[Bartalotti {\normalfont et~al.}(2021)Bartalotti, Brummet and
  Dieterle]{Bartalotti-Brummet-Dieterle_2021_JBES}
Bartalotti, O., Brummet, Q., and Dieterle, S. (2021), \enquote{A Correction for
  Regression Discontinuity Designs with Group-specific Mismeasurement of the
  Running Variable,} \emph{Journal of Business \& Economic Statistics}, 39,
  833--848.
\bibitem[Boon {\normalfont et~al.}(2021)Boon, Craig, Thomson, Campbell and
  Moore]{boon2021regression}
Boon, M.~H., Craig, P., Thomson, H., Campbell, M., and Moore, L. (2021),
  \enquote{Regression discontinuity designs in health: a systematic review,}
  \emph{Epidemiology}, 32, 87.
\bibitem[Bor {\normalfont et~al.}(2017)Bor, Fox, Rosen, Venkataramani, Tanser,
  Pillay and B{\"a}rnighausen]{Bor-Fox-Rosen-etal_2017_PloSMed}
Bor, J., Fox, M.~P., Rosen, S., Venkataramani, A., Tanser, F., Pillay, D., and
  B{\"a}rnighausen, T. (2017), \enquote{Treatment eligibility and retention in
  clinical HIV care: A regression discontinuity study in South Africa,}
  \emph{PLoS Medicine}, 14, e1002463.
\bibitem[Bor {\normalfont et~al.}(2015)Bor, Moscoe and
  B{\"a}rnighausen]{Bor-Moscoe-Barnighausen_2015_Epidem}
Bor, J., Moscoe, E., and B{\"a}rnighausen, T. (2015), \enquote{Three approaches
  to causal inference in regression discontinuity designs,}
  \emph{Epidemiology}, 26, e28--e30.
\bibitem[Bor {\normalfont et~al.}(2014)Bor, Moscoe, Mutevedzi, Newell and
  B{\"a}rnighausen]{Bor-Moscoe-Mutevedzi-Newell-Barnighausen_2014_Epidem}
Bor, J., Moscoe, E., Mutevedzi, P., Newell, M.-L., and B{\"a}rnighausen, T.
  (2014), \enquote{Regression Discontinuity Designs in Epidemiology: Causal
  Inference without Randomized Trials,} \emph{Epidemiology}, 25, 729--737.
\bibitem[Branson and Keele(2020)]{branson2020evaluating}
Branson, Z., and Keele, L. (2020), \enquote{Evaluating a key instrumental
  variable assumption using randomization tests,} \emph{American Journal of
  Epidemiology}, 189, 1412--1420.
\bibitem[Calonico {\normalfont et~al.}(2018)Calonico, Cattaneo and
  Farrell]{Calonico-Cattaneo-Farrell_2018_JASA}
Calonico, S., Cattaneo, M.~D., and Farrell, M.~H. (2018), \enquote{On the
  Effect of Bias Estimation on Coverage Accuracy in Nonparametric Inference,}
  \emph{Journal of the American Statistical Association}, 113, 767--779.
\bibitem[Calonico {\normalfont et~al.}(2020)Calonico, Cattaneo and
  Farrell]{Calonico-Cattaneo-Farrell_2020_ECTJ}
\leavevmode\vrule height .65ex depth -.6ex width 3em\  (2020), \enquote{Optimal
  Bandwidth Choice for Robust Bias Corrected Inference in Regression
  Discontinuity Designs,} \emph{Econometrics Journal}, 23, 192--210.
\bibitem[Calonico {\normalfont et~al.}(2022)Calonico, Cattaneo and
  Farrell]{Calonico-Cattaneo-Farrell_2022_Bernoulli}
\leavevmode\vrule height .65ex depth -.6ex width 3em\  (2022),
  \enquote{Coverage Error Optimal Confidence Intervals for Local Polynomial
  Regression,} \emph{Bernoulli}, 28, 2998--3022.
\bibitem[Calonico {\normalfont et~al.}(2019)Calonico, Cattaneo, Farrell and
  Titiunik]{Calonico-Cattaneo-Farrell-Titiunik_2019_RESTAT}
Calonico, S., Cattaneo, M.~D., Farrell, M.~H., and Titiunik, R. (2019),
  \enquote{Regression Discontinuity Designs using Covariates,} \emph{Review of
  Economics and Statistics}, 101, 442--451.
\bibitem[Calonico {\normalfont et~al.}(2014)Calonico, Cattaneo and
  Titiunik]{Calonico-Cattaneo-Titiunik_2014_ECMA}
Calonico, S., Cattaneo, M.~D., and Titiunik, R. (2014), \enquote{Robust
  Nonparametric Confidence Intervals for Regression-Discontinuity Designs,}
  \emph{Econometrica}, 82, 2295--2326.
\bibitem[Calonico {\normalfont et~al.}(2015)Calonico, Cattaneo and
  Titiunik]{Calonico-Cattaneo-Titiunik_2015_JASA}
\leavevmode\vrule height .65ex depth -.6ex width 3em\  (2015), \enquote{Optimal
  Data-Driven Regression Discontinuity Plots,} \emph{Journal of the American
  Statistical Association}, 110, 1753--1769.
\bibitem[Card {\normalfont et~al.}(2015)Card, Lee, Pei and
  Weber]{Card-Lee-Pei-Weber_2015_ECMA}
Card, D., Lee, D.~S., Pei, Z., and Weber, A. (2015), \enquote{Inference on
  Causal Effects in a Generalized Regression Kink Design,} \emph{Econometrica},
  83, 2453--2483.
\bibitem[Cattaneo {\normalfont et~al.}(2023{\natexlab{a}})Cattaneo, Crump,
  Farrell and Feng]{Cattaneo-Crump-Farrell-Feng_2023_Binscatter}
Cattaneo, M.~D., Crump, R.~K., Farrell, M.~H., and Feng, Y.
  (2023{\natexlab{a}}), \enquote{On Binscatter,}
  \emph{\emph{arXiv:1902.09608}}.
\bibitem[Cattaneo {\normalfont et~al.}(2015)Cattaneo, Frandsen and
  Titiunik]{Cattaneo-Frandsen-Titiunik_2015_JCI}
Cattaneo, M.~D., Frandsen, B., and Titiunik, R. (2015), \enquote{Randomization
  Inference in the Regression Discontinuity Design: An Application to Party
  Advantages in the U.S. Senate,} \emph{Journal of Causal Inference}, 3, 1--24.
\bibitem[Cattaneo {\normalfont et~al.}(2020{\natexlab{a}})Cattaneo, Idrobo and
  Titiunik]{Cattaneo-Idrobo-Titiunik_2020_Book}
Cattaneo, M.~D., Idrobo, N., and Titiunik, R. (2020{\natexlab{a}}), \emph{A
  Practical Introduction to Regression Discontinuity Designs: Foundations},
  Cambridge Elements: Quantitative and Computational Methods for Social
  Science, Cambridge University Press.
\bibitem[Cattaneo {\normalfont et~al.}(2023{\natexlab{b}})Cattaneo, Idrobo and
  Titiunik]{Cattaneo-Idrobo-Titiunik_2023_Book}
\leavevmode\vrule height .65ex depth -.6ex width 3em\  (2023{\natexlab{b}}),
  \emph{A Practical Introduction to Regression Discontinuity Designs:
  Extensions}, Cambridge Elements: Quantitative and Computational Methods for
  Social Science, Cambridge University Press.
\bibitem[Cattaneo {\normalfont et~al.}(2020{\natexlab{b}})Cattaneo, Jansson and
  Ma]{Cattaneo-Jansson-Ma_2020_JASA}
Cattaneo, M.~D., Jansson, M., and Ma, X. (2020{\natexlab{b}}), \enquote{Simple
  Local Polynomial Density Estimators,} \emph{Journal of the American
  Statistical Association}, 115, 1449--1455.
\bibitem[Cattaneo {\normalfont et~al.}(2023{\natexlab{c}})Cattaneo, Keele and
  Titiunik]{Cattaneo-Keele-Titiunik_2023_HandbookCh}
Cattaneo, M.~D., Keele, L., and Titiunik, R. (2023{\natexlab{c}}),
  \enquote{Covariate Adjustment in Regression Discontinuity Designs,} in
  \emph{Handbook of Matching and Weighting in Causal Inference}, eds. D.~S.~S.
  J.~R.~Zubizarreta, E. A.~Stuart and P.~R. Rosenbaum, chapter~8, Chapman \&
  Hall, pp.\  153--168.
\bibitem[Cattaneo {\normalfont et~al.}(2016)Cattaneo, Keele, Titiunik and
  Vazquez-Bare]{Cattaneo-Keele-Titiunik-VazquezBare_2016_JOP}
Cattaneo, M.~D., Keele, L., Titiunik, R., and Vazquez-Bare, G. (2016),
  \enquote{Interpreting Regression Discontinuity Designs with Multiple
  Cutoffs,} \emph{Journal of Politics}, 78, 1229--1248.
\bibitem[Cattaneo {\normalfont et~al.}(2021)Cattaneo, Keele, Titiunik and
  Vazquez-Bare]{Cattaneo-Keele-Titiunik-VazquezBare_2021_JASA}
\leavevmode\vrule height .65ex depth -.6ex width 3em\  (2021),
  \enquote{Extrapolating Treatment Effects in Multi-Cutoff Regression
  Discontinuity Designs,} \emph{Journal of the American Statistical
  Association}, 116, 1941--1952.
\bibitem[Cattaneo and Titiunik(2022)]{Cattaneo-Titiunik_2022_ARE}
Cattaneo, M.~D., and Titiunik, R. (2022), \enquote{Regression Discontinuity
  Designs,} \emph{Annual Review of Economics}, 14, 821--851.
\bibitem[Cattaneo {\normalfont et~al.}(2017)Cattaneo, Titiunik and
  Vazquez-Bare]{Cattaneo-Titiunik-VazquezBare_2017_JPAM}
Cattaneo, M.~D., Titiunik, R., and Vazquez-Bare, G. (2017), \enquote{Comparing
  Inference Approaches for RD Designs: A Reexamination of the Effect of Head
  Start on Child Mortality,} \emph{Journal of Policy Analysis and Management},
  36, 643--681.
\bibitem[Cattaneo {\normalfont et~al.}(2023{\natexlab{d}})Cattaneo, Titiunik
  and Yu]{Cattaneo-Titiunik-Yu_2023_wp}
Cattaneo, M.~D., Titiunik, R., and Yu, R. (2023{\natexlab{d}}),
  \enquote{Estimation and Inference in Boundary Discontinuity Designs,}
  \emph{working paper}.
\bibitem[Cattaneo and Vazquez-Bare(2016)]{Cattaneo-VazquezBare_2016_ObsStud}
Cattaneo, M.~D., and Vazquez-Bare, G. (2016), \enquote{The Choice of
  Neighborhood in Regression Discontinuity Designs,} \emph{Observational
  Studies}, 2, 134--146.
\bibitem[Craig {\normalfont et~al.}(2017)Craig, Katikireddi, Leyland and
  Popham]{Craig-Katikireddi-Leyland-Popham_2017_ARPH}
Craig, P., Katikireddi, S.~V., Leyland, A., and Popham, F. (2017),
  \enquote{Natural Experiments: An Overview of Methods, Approaches, and
  Contributions to Public Health Intervention Research,} \emph{Annual Review of
  Public Health}, 38, 39--56.
\bibitem[Davies {\normalfont et~al.}(2017)Davies, Thomas, Taylor, Taylor,
  Martin, Munaf{\`o} and Windmeijer]{Davies-etal_2017_IJE}
Davies, N.~M., Thomas, K.~H., Taylor, A.~E., Taylor, G.~M., Martin, R.~M.,
  Munaf{\`o}, M.~R., and Windmeijer, F. (2017), \enquote{How to compare
  instrumental variable and conventional regression analyses using negative
  controls and bias plots,} \emph{International Journal of Epidemiology}, 46,
  2067--2077.
\bibitem[De~Magalh{\~a}es {\normalfont et~al.}(2020)De~Magalh{\~a}es,
  Hangartner, Hirvonen, Meril{\"a}inen, Ruiz and
  Tukiainen]{DeMagalhaes-etal_2020_wp}
De~Magalh{\~a}es, L., Hangartner, D., Hirvonen, S., Meril{\"a}inen, J., Ruiz,
  N., and Tukiainen, J. (2020), \enquote{How Much Should We Trust Regression
  Discontinuity Design Estimates? Evidence from Experimental Benchmarks of the
  Incumbency Advantage,} \emph{working paper}.
\bibitem[Dong(2018)]{Dong_2018_OBES}
Dong, Y. (2018), \enquote{Alternative Assumptions to Identify LATE in Fuzzy
  Regression Discontinuity Designs,} \emph{Oxford Bulletin of Economics and
  Statistics}, 80, 1020--1027.
\bibitem[Dong {\normalfont et~al.}(2021)Dong, Lee and
  Gou]{Dong-Lee-Gou_2021_JASA}
Dong, Y., Lee, Y.-Y., and Gou, M. (2021), \enquote{Regression Discontinuity
  Designs with a Continuous Treatment,} \emph{Journal of the American
  Statistical Association}, 118, 208--221.
\bibitem[Ernst(2004)]{Ernst_2004_SS}
Ernst, M.~D. (2004), \enquote{Permutation Methods: A Basis for Exact
  Inference,} \emph{Statistical Science}, 19, 676--685.
\bibitem[Fan and Gijbels(1996)]{Fan-Gijbels_1996_Book}
Fan, J., and Gijbels, I. (1996), \emph{Local polynomial Modelling and Its
  Applications}, Vol.~66, CRC Press.
\bibitem[Feir {\normalfont et~al.}(2016)Feir, Lemieux and
  Marmer]{Feir-Lemieux-Marmer_2016_JBES}
Feir, D., Lemieux, T., and Marmer, V. (2016), \enquote{Weak identification in
  fuzzy regression discontinuity designs,} \emph{Journal of Business \&
  Economic Statistics}, 34, 185--196.
\bibitem[Ganong and J{\"a}ger(2018)]{Ganong-Jager_2018_JASA}
Ganong, P., and J{\"a}ger, S. (2018), \enquote{A Permutation Test for the
  Regression Kink Design,} \emph{Journal of the American Statistical
  Association}, 113, 494--504.
\bibitem[Glymour {\normalfont et~al.}(2012)Glymour, Tchetgen~Tchetgen and
  Robins]{Glymour-TchetgenTchetgen-Robins_2012_AJE}
Glymour, M.~M., Tchetgen~Tchetgen, E.~J., and Robins, J.~M. (2012),
  \enquote{Credible Mendelian randomization studies: {A}pproaches for
  evaluating the instrumental variable assumptions,} \emph{American Journal of
  Epidemiology}, 175, 332--339.
\bibitem[Hahn {\normalfont et~al.}(2001)Hahn, Todd and van~der
  Klaauw]{Hahn-Todd-vanderKlaauw_2001_ECMA}
Hahn, J., Todd, P., and van~der Klaauw, W. (2001), \enquote{Identification and
  Estimation of Treatment Effects with a Regression-Discontinuity Design,}
  \emph{Econometrica}, 69, 201--209.
\bibitem[Han {\normalfont et~al.}(2020)Han, Lien and Yang]{han2020patient}
Han, H.-W., Lien, H.-M., and Yang, T.-T. (2020), \enquote{Patient Cost-Sharing
  and Healthcare Utilization in Early Childhood: Evidence from a Regression
  Discontinuity Design,} \emph{American Economic Journal: Economic Policy}, 12,
  238--78.
\bibitem[Hern{\'a}n(2018)]{Hernan_2018_AJPH}
Hern{\'a}n, M.~A. (2018), \enquote{The C-word: Scientific Euphemisms Do Not
  Improve Causal Inference from Observational Data,} \emph{American Journal of
  Public Health}, 108, 616--619.
\bibitem[Hern\'an and Robins(2022)]{Hernan-Robins_2022_Book}
Hern\'an, M.~A., and Robins, J.~M. (2022), \emph{Causal Inference: What If},
  Boca Raton: Chapman \& Hall/CRC.
\bibitem[Houlihan {\normalfont et~al.}(2010)Houlihan, Bland, Mutevedzi,
  Lessells, Ndirangu, Thulare and Newell]{Houlihan-etal_2010_IJE}
Houlihan, C.~F., Bland, R.~M., Mutevedzi, P.~C., Lessells, R.~J., Ndirangu, J.,
  Thulare, H., and Newell, M.-L. (2010), \enquote{Cohort Profile: Hlabisa HIV
  Treatment and Care Programme,} \emph{International Journal of Epidemiology},
  40, 318--326.
\bibitem[Hyytinen {\normalfont et~al.}(2018)Hyytinen, Meril\"ainen, Saarimaa,
  Toivanen and Tukiainen]{Hyytinen-etal_2018_QE}
Hyytinen, A., Meril\"ainen, J., Saarimaa, T., Toivanen, O., and Tukiainen, J.
  (2018), \enquote{When Does Regression Discontinuity Design Work? Evidence
  from Random Election Outcomes,} \emph{Quantitative Economics}, 9, 1019--1051.
\bibitem[Imbens and Rubin(2015)]{Imbens-Rubin_2015_Book}
Imbens, G.~W., and Rubin, D.~B. (2015), \emph{Causal Inference in Statistics,
  Social, and Biomedical Sciences}, Cambridge University Press.
\bibitem[Kamat(2018)]{Kamat_2018_ET}
Kamat, V. (2018), \enquote{On Nonparametric Inference in the Regression
  Discontinuity Design,} \emph{Econometric Theory}, 34, 694--703.
\bibitem[Kang {\normalfont et~al.}(2018)Kang, Peck and
  Keele]{Kang-Peck-Keele_2018_JRSSA}
Kang, H., Peck, L., and Keele, L. (2018), \enquote{Inference for Instrumental
  Variables: A Randomization Inference Approach,} \emph{Journal of The Royal
  Statistical Society, Series A}, 181, 1231--1254.
\bibitem[Kaptchuk and Miller(2015)]{KaptchukMiller2015-NEJM}
Kaptchuk, T.~J., and Miller, F.~G. (2015), \enquote{Placebo effects in
  medicine,} \emph{N Engl J Med}, 373, 8--9.
\bibitem[Keele {\normalfont et~al.}(2017)Keele, Small and
  Grieve]{Keele-Small-Grieve_2017_JRSSA}
Keele, L.~J., Small, D.~S., and Grieve, R. (2017), \enquote{Randomization Based
  Instrumental Variables Methods for Binary Outcomes with an Application to the
  IMPROVE Trial,} \emph{Journal of The Royal Statistical Society, Series A},
  180, 569--586.
\bibitem[Keele and Titiunik(2015)]{Keele-Titiunik_2015_PA}
Keele, L.~J., and Titiunik, R. (2015), \enquote{Geographic Boundaries as
  Regression Discontinuities,} \emph{Political Analysis}, 23, 127--155.
\bibitem[Keele {\normalfont et~al.}(2015)Keele, Titiunik and
  Zubizarreta]{Keele-Titiunik-Zubizarreta_2015_JRSSA}
Keele, L.~J., Titiunik, R., and Zubizarreta, J. (2015), \enquote{Enhancing a
  Geographic Regression Discontinuity Design Through Matching to Estimate the
  Effect of Ballot Initiatives on Voter Turnout,} \emph{Journal of the Royal
  Statistical Society: Series A}, 178, 223--239.
\bibitem[Keele {\normalfont et~al.}(2019)Keele, Zhao, Kelz and
  Small]{Keele-Zhao-Kelz-Small_2018_MC}
Keele, L.~J., Zhao, Q., Kelz, R.~R., and Small, D.~S. (2019),
  \enquote{Falsification Tests for Instrumental Variable Designs with an
  Application to Tendency to Operate,} \emph{Medical Care}, 57, 167--171.
\bibitem[Korting {\normalfont et~al.}(2023)Korting, Lieberman, Matsudaira, Pei
  and Shen]{Korting-etal_2023_visual}
Korting, C., Lieberman, C., Matsudaira, J., Pei, Z., and Shen, Y. (2023),
  \enquote{Visual Inference and Graphical Representation in Regression
  Discontinuity Designs,} \emph{Quartely Journal of Economics}.
\bibitem[Maciejewski and Basu(2020)]{Maciejewski-Basu_2020_JAMA}
Maciejewski, M.~L., and Basu, A. (2020), \enquote{Regression Discontinuity
  Design,} \emph{JAMA}, 324, 381--382.
\bibitem[McCrary(2008)]{McCrary_2008_JoE}
McCrary, J. (2008), \enquote{Manipulation of the running variable in the
  regression discontinuity design: A density test,} \emph{Journal of
  Econometrics}, 142, 698--714.
\bibitem[O'Keeffe {\normalfont et~al.}(2014)O'Keeffe, Geneletti, Baio,
  Sharples, Nazareth and
  Petersen]{OKeeffe-Geneletti-Baio-sharples-Nazareth-Petersen_2014_BMJ}
O'Keeffe, A.~G., Geneletti, S., Baio, G., Sharples, L.~D., Nazareth, I., and
  Petersen, I. (2014), \enquote{Regression Discontinuity Designs: An Approach
  to the Evaluation of Treatment Efficacy in Primary Care Using Observational
  Data,} \emph{BMJ}, 349:g5293.
\bibitem[Paik {\normalfont et~al.}(2006)Paik, Tang, Shak, Kim, Baker, Kim,
  Cronin, Baehner, Watson, Bryant, Costantino, Geyer, Wickerham and
  Wolmark]{Paik-etal_2006_JCO}
Paik, S., Tang, G., Shak, S., Kim, C., Baker, J., Kim, W., Cronin, M., Baehner,
  F.~L., Watson, D., Bryant, J., Costantino, J.~P., Geyer, C. E.~J., Wickerham,
  D.~L., and Wolmark, N. (2006), \enquote{Gene Expression and Benefit of
  Chemotherapy in Women With Node-Negative, Estrogen Receptor--Positive Breast
  Cancer,} \emph{Journal of Clinical Oncology}, 24, 3726--3734.
\bibitem[Papay {\normalfont et~al.}(2011)Papay, Willett and
  Murnane]{PapayWillettMurnane2011-JoE}
Papay, J.~P., Willett, J.~B., and Murnane, R.~J. (2011), \enquote{Extending the
  regression-discontinuity approach to multiple assignment variables,}
  \emph{Journal of Econometrics}, 161, 203--207.
\bibitem[Pizer(2016)]{Pizer_2016_HSR}
Pizer, S.~D. (2016), \enquote{Falsification testing of instrumental variables
  methods for comparative effectiveness research,} \emph{Health Services
  Research}, 51, 790--811.
\bibitem[Reardon and Robinson(2012)]{ReardonRobinson2012-JREE}
Reardon, S.~F., and Robinson, J.~P. (2012), \enquote{Regression discontinuity
  designs with multiple rating-score variables,} \emph{Journal of Research on
  Educational Effectiveness}, 5, 83--104.
\bibitem[Rosenbaum(2010)]{Rosenbaum_2010_Book}
Rosenbaum, P.~R. (2010), \emph{Design of Observational Studies}, Springer.
\bibitem[Sekhon and Titiunik(2016)]{Sekhon-Titiunik_2016_ObsStud}
Sekhon, J.~S., and Titiunik, R. (2016), \enquote{Understanding Regression
  Discontinuity Designs as Observational Studies,} \emph{Observational
  Studies}, 2, 174--182.
\bibitem[Sekhon and Titiunik(2017)]{Sekhon-Titiunik_2017_AIE}
\leavevmode\vrule height .65ex depth -.6ex width 3em\  (2017), \enquote{On
  Interpreting the Regression Discontinuity Design as a Local Experiment,} in
  \emph{Regression Discontinuity Designs: Theory and Applications (Advances in
  Econometrics, volume 38)}, eds. M.~D. Cattaneo and J.~C. Escanciano, Emerald
  Group Publishing, pp.\  1--28.
\bibitem[Sparano and Paik(2008)]{Sparano-Paik_2008_JCO}
Sparano, J.~A., and Paik, S. (2008), \enquote{Development of the 21-Gene Assay
  and Its Application in Clinical Practice and Clinical Trials,} \emph{Journal
  of Clinical Oncology}, 26, 721--728.
\bibitem[Tanser {\normalfont et~al.}(2007)Tanser, Hosegood, B{\"a}rnighausen,
  Herbst, Nyirenda, Muhwava, Newell, Viljoen, Mutevedzi and
  Newell]{Tanser-etal_2007_IJE}
Tanser, F., Hosegood, V., B{\"a}rnighausen, T., Herbst, K., Nyirenda, M.,
  Muhwava, W., Newell, C., Viljoen, J., Mutevedzi, T., and Newell, M.-L.
  (2007), \enquote{Cohort Profile: Africa Centre Demographic Information System
  (ACDIS) and Population-based HIV Survey,} \emph{International Journal of
  Epidemiology}, 37, 956--962.
\bibitem[Thistlethwaite and Campbell(1960)]{Thistlethwaite-Campbell_1960_JEP}
Thistlethwaite, D.~L., and Campbell, D.~T. (1960),
  \enquote{Regression-Discontinuity Analysis: An Alternative to the Ex-Post
  Facto Experiment,} \emph{Journal of Educational Psychology}, 51, 309--317.
\bibitem[Titiunik(2021)]{Titiunik_2021_HandbookCh}
Titiunik, R. (2021), \enquote{Natural Experiments,} in \emph{Advances in
  Experimental Political Science}, eds. J.~N. Druckman and D.~P. Gree,
  chapter~6, Cambridge University Press, pp.\  103--129.
\bibitem[Tuvaandorj(2020)]{Tuvaandorj_2020_JoE}
Tuvaandorj, P. (2020), \enquote{Regression Discontinuity Designs, White Noise
  Models, and Minimax,} \emph{Journal of Econometrics}, 218, 587--608.
\bibitem[Xu(2017)]{Xu_2017_JoE}
Xu, K.-L. (2017), \enquote{Regression Discontinuity with Categorical Outcomes,}
  \emph{Journal of Econometrics}, 201, 1--18.
\end{thebibliography}

\clearpage

\appendix

\section{Annotated Sample Code}

Full replication files can be found at \url{https://rdpackages.github.io/}.  The replication files include the data and scripts for \texttt{Python}, \texttt{R} and \texttt{Stata} to generate every result in the paper. Below we provide an abbreviated version of \texttt{R} code for the ART application for illustration purposes. Note that these commands will not exactly replicate the results in the tables due to handling of missing values. For exact replication of tables, see the replication codes.

\small
\begin{verbatim}
> library(foreign)
> library(rdrobust)
> library(rddensity)
> library(rdlocrand)
> 
> # Read dataset
> data <- read.dta("CKT_2023_SIM--ART.dta")
> X = data$cd4 
> Y = data$visit_test_6_18
> D = data$art_6m
 
> ## Generate RD Plot
> rdplot(Y, X, c=350, y.label="Retained", x.label="CD4 Count", p=3)

> ## Continuity-Based RD Analysis: ITT effect on outcome
> rdrobust(Y, X, c=350)
> ## Continuity-Based RD Analysis: ITT effect on treatment
> rdrobust(D, X, c=350)
> ## Continuity-Based RD Analysis: Fuzzy effect on outcome
> rdrobust(Y, X, c=350, fuzzy=D)

> ## Local Randomization Window selection
> vars <- c("age1", "age2", "age3", "age4","age5", "age6", "age7", "age8", "qtr1", "qtr2",
+           "qtr3", "qtr4", "qtr5", "qtr6", "clinic_a", "clinic_b", "clinic_c")
> Z <- data[vars]
> out = rdwinselect(X, Z, c=350, seed = 50, reps = 1000, wstep=1)

> ## Local Randomization Analysis: ITT effect on outcome
> ci_vec = c(0.05, seq(from = -.5, to = .5, by = 0.01))
> rdrandinf(Y, X, cutoff = 350, seed= 5023, wl=346, wr=354, ci=ci_vec)  

> ## Local Randomization Analysis: Fuzzy effect on outcome
> ci_vec = c(0.05, seq(from = -1, to = 1, by = 0.01))
> rdrandinf(Y, X, fuzzy=c(fuzzy.tr=D, fuzzy.stat="tsls"),  
+             cutoff = 350, seed= 5023, wl=346, wr=354, ci=ci_vec)
\end{verbatim}

\end{document}